%% file: main.tex
\input{sections/preamble}

\input{sections/title}

\begin{document}
\label{firstpage}
\pagerange{\pageref{firstpage}--\pageref{lastpage}}
\maketitle

\input{sections/abstract}

\begin{keywords}
ISM: supernova remnants -- catalogues
\end{keywords}



\input{sections/introduction}
\input{sections/section_2}
\input{sections/section_3}
\input{sections/results}
\input{sections/discussion}
\input{sections/conclusions}






\section*{Acknowledgements}

We thank Prof. Miroslav Filipovic for his thoughtful and constructive suggestions which greatly enhanced the quality of our manuscript. This work was supported by the \emph{Found\-ation for Re\-search \& Tech\-nology – He\-llas} (FORTH). IL and AZ  acknowledge support from the \emph{FORTH Synergy Grant PARSEC}. AZ  acknowledges support from the European Union’s Horizon 2020 research and innovation programme under the \emph{Marie Skłodowska-Curie RISE} action, grant agreement No 873089 (ASTROSTAT-II). MK acknowledges financial support from MICINN (Spain) through the programme \emph{Juan de la Cierva-Incorporación [JC2022-049447-I]} and from AGAUR, CSIC, MCIN and AEI 10.13039/501100011033 under projects PID2023-151307NB-I00, PIE 20215AT016, CEX2020-001058-M, and 2021-SGR-01270.

\section*{Data Availability}

All the tabular data presented in appendices A through C are available on the CDS.



\bibliographystyle{mnras}
\bibliography{references}


\onecolumn
\appendix

\begin{landscape}
\section{The sample}
\label{app:objects}
\input{tables/table}
\end{landscape}

\newpage
\begin{landscape}
\section{The dataset}
\label{app:dataset}
\input{appendices/appendix_b}
\input{tables/dataset}
\end{landscape}

\newpage
\begin{landscape}
\section{Line strengths}
\label{app:lines}
\input{tables/line_table}


\bsp	
\label{lastpage}
\end{landscape}
\end{document}

%% file: sections/preamble.tex
\documentclass[fleqn,usenatbib]{mnras}

\usepackage{newtxtext,newtxmath}

\usepackage[T1]{fontenc}

\DeclareRobustCommand{\VAN}[3]{#2}
\let\VANthebibliography\thebibliography
\def\thebibliography{\DeclareRobustCommand{\VAN}[3]{##3}\VANthebibliography}

\DeclareUnicodeCharacter{0301}{\'{}}
\DeclareUnicodeCharacter{2212}{-}
\DeclareUnicodeCharacter{2272}{\ensuremath{\leqslant}} 
\DeclareUnicodeCharacter{2248}{\ensuremath{\approx}}   
\DeclareUnicodeCharacter{2273}{\ensuremath{\geqslant}} 

\usepackage{graphicx}	
\usepackage{amsmath}	
\usepackage{xcolor}
\usepackage{lipsum}
\usepackage{float}
\usepackage{rotating}
\usepackage{longtable}
\usepackage{caption}
\usepackage{ragged2e}
\usepackage{pdflscape}
\usepackage{xltabular}
\usepackage[normalem]{ulem}
\useunder{\uline}{\ul}{}
\usepackage{lscape}
\usepackage{orcidlink}

\usepackage{hyperref}
\hypersetup{colorlinks=true, 
            citecolor=blue, 
            linkcolor=blue, 
            filecolor=magenta, 
            urlcolor=cyan,
            pdftitle={A systematic meta-analysis of physical parameters of Galactic supernova remnants},
            pdfpagemode=FullScreen}



%% file: sections/title.tex
\title[Study of physical parameters of Galactic SNRs]{A systematic meta-analysis of physical parameters of Galactic supernova remnants}

\author[I. Chousein-Basia et al.]{
I. Chousein-Basia$^{1,2,3}$\thanks{E-mail: i.chousein.basia@ia.forth.gr}\orcidlink{0009-0008-3588-2388},
A. Zezas$^{1,2}$\thanks{E-mail: azezas@physics.uoc.gr}\orcidlink{0000-0001-8952-676X},
I. Leonidaki$^{1,2}$\orcidlink{0000-0001-5547-3401}
and M. Kopsacheili$^{4,5}$\orcidlink{0000-0002-3563-819X}
\\
$^{1}$Department of Physics, University of Crete, Voutes University Campus, Iraklio 71003, Greece\\
$^{2}$Institute of Astrophysics, Foundation for Research and Technology-Hellas (FORTH), Iraklio 71110, Greece\\
$^{3}$Faculty of Science and Engineering, University of Groningen, Groningen 9747 AG, Netherlands\\
$^{4}$Institute of Space Sciences (ICE, CSIC), Campus UAB, Carrer de Can Magrans-2097, s/n, E-08193 Barcelona, Spain\\
$^{5}$Institut d’Estudis Espacials de Catalunya (IEEC), 08860 Castelldefels-2099 (Barcelona), Spain
}

\date{Accepted XXX. Received YYY; in original form ZZZ}

\pubyear{\the\year{}}

%% file: sections/abstract.tex
\begin{abstract}
Supernova remnants (SNRs) are the aftermath of massive stellar explosions or of a white dwarf in a binary system, representing critical phases in the life cycle of stars and playing an important role in galactic evolution. Physical properties of SNRs
such as their shock velocity, density and age are important elements for constraining models for their evolution and understanding the physical processes responsible for their morphological appearance and emission processes. Our study provides, for the first time, a comprehensive statistical analysis of the physical parameters in 64 Galactic SNRs both as a population as well as regions within individual objects. These 64 objects represent the subset of the 310 known Galactic SNRs for which there are published optical data, from which we compiled their physical parameters through an exhaustive literature survey. Through a systematic statistical analysis accounting for uncertainties and/or upper and lower limits in these parameters we obtain distributions of the electron density and shock velocity in the studied SNRs and regions within them. This information is combined with constraints on their age and type. Analysis of electron density and shock velocity distributions for the entire sample of SNRs shows that they are consistent with a log-normal distribution and a skewed log-normal distribution, respectively. Within individual remnants, our study reveals that electron density and shock velocity show larger scatter in younger objects, reflecting the varying conditions of the ambient medium immediately surrounding the explosion epicenter and their impact on SNR evolution. Comparison of the dependence of the shock velocity and density on the supernova age with expectations from theoretical models shows good agreement.
\end{abstract}

%% file: sections/introduction.tex
\section{Introduction}
Supernova explosions are one of the most energetic phenomena ob\-served in the universe. These events signify the dramatic conclusion of a massive star's or an accreting white dwarf's life cycle. During a supernova explosion, a shock wave is produced, propelling stellar material outward into the interstellar medium (ISM) at supersonic speeds. This shock wave causes the ISM to condense and heat up, resulting in the emission of radiation across a significant portion of the electromag\-netic spectrum. Observing and studying supernova remants (SNRs) provides valuable insights into the life cycles of stars, the dynamics of interstellar matter, and nucleosynthesis. They also serve as laboratories for understanding shock physics, particle acceleration, and magnetic field amplification in extreme environments \citep[e.g.,][]{Morlino_2013, Caprioli_2023}.

Surveys of large numbers of both Galactic and extragalactic SNRs allow us to obtain information on the physical properties of these objects, such as their temperature, electron density and shock velocity distributions. There are some advantages and disadvantages in each case. Extinction from interstellar material would make observations of SNRs within our Galaxy challenging. On the other hand, owing to their proximity, 
Galactic SNRs allow us to obtain high precision, spatially resolved, measurements on their physical properties such as  electron density, temperature, shock and expansion velocity, as well as their morphology and environment \citep[e.g.,][]{1983MNRAS.204..273L, 2001MNRAS.325..287W, 2013Sci...340...45N}.

These measurements have provided valuable information that informed models for SNR evolution \citep[e.g.,][]{2018ApJ...866....9L, Leahy_2020}. On the other hand, observations of extragalactic SNRs give us a more representative picture of their populations in different environments \citep[e.g.,][]{2010ApJ...725..842L, 2013MNRAS.429..189L, Long_2017, Kopsacheili_2021}, they overcome the problem of distance determination, and they are not subject to the large extinction column densities in our Galaxy's disk. This comes at the cost of lower spatial resolution that does not allow detailed studies of their morphology and physical parameters, with the only exception of the Magellanic Clouds \citep[e.g.,][]{2016A&A...585A.162M, Maggi_2019, 2022MNRAS.515.4099K, Alsaberi_2023}.

To date, no systematic analysis has been conducted on the {\textit{physical}} properties of Galactic SNRs, in contrast to several such analyses performed on the \textit{observational} properties of Galactic SNRs such as angular size and luminosity, and other properties such as age and distance \citep[e.g.,][]{2005ChJAA...5..165X, 2023ApJS..265...53R, 2024sros.confE..17G} or the properties of extragalactic SNRs \citep[e.g.,][]{1997ApJS..112...49M, 2010ApJS..187..495L, 2010ApJ...725..842L, 2013MNRAS.429..189L, 2017hsn..book.2005L, 2023MNRAS.518.2574B, 2024MNRAS.530.1078K}. 

Thanks to a large body of works presenting information on individual objects within our Galaxy or regions within them, we are able to perform a systematic meta-analysis of these data within a statistical framework in order to obtain a picture of the overall trends of the physical properties in the population of SNRs within our Galaxy. These can then be used for systematic comparisons with extragalactic SNR populations in different environments and with SNR evolution models.

In this study, we focus on physical parameters derived from optical observations. In Section \ref{sec:2}, we describe the sample selection and data aggregation. In Section \ref{sec:3}, we present the methods for the homogenization of the data, and in Section \ref{sec:4} we present our results, including the statistical analysis of the SNR physical parameters. In Section \ref{sec:discussion}, we discuss the results from the statistical analysis in the context of SNR evolution models, and in Section \ref{sec:conclusions} we summarize the main results from this work.

%% file: sections/section_2.tex
\section{Sample selection \& data consolidation}
\label{sec:2}
    The sample was selected based on the Galactic SNR catalog of \cite{Green}. We thoroughly examined all publications available in the literature for all 310 SNRs listed in the catalog. Due to a lack of systematic analyses of physical parameters derived from optical observations, we focused on studies based on optical observations. Studies in other wavelengths have been very successful in constructing detailed morphological maps of SNRs and obtaining spatially resolved measurements of physical parameters, such as electron density, temperature, and velocity structures within them (see, e.g., \citealt{Yan_2020}). X-rays, visible only in the early evolution of SNRs, can provide valuable insights into their inner structure thanks to the emission of their shocked ejecta. Radio synchrotron emission is direct evidence of particle acceleration at shock fronts or relativistic particles in pulsar-wind nebulae  while radio absorption features imprinted on synchrotron continuum (such as HI 21 cm and molecular lines) can yield information on ISM composition in the vicinity of the remnant.
    
    However, optical studies have the advantage of allowing us to explore the outer regions, where knotty or filamentary structures give rise to most of the optical emission we observe. Additionally, optical studies are important because they provide direct information on the heated gas and shock over several evolutionary phases of the SNR. When SNRs emit strongly in the visible part of the electromagnetic spectrum, they are typically at the end of their adiabatic phase or at their radiative phase of evolution.

    \begin{figure}
    \centering
    \includegraphics[width=\columnwidth]{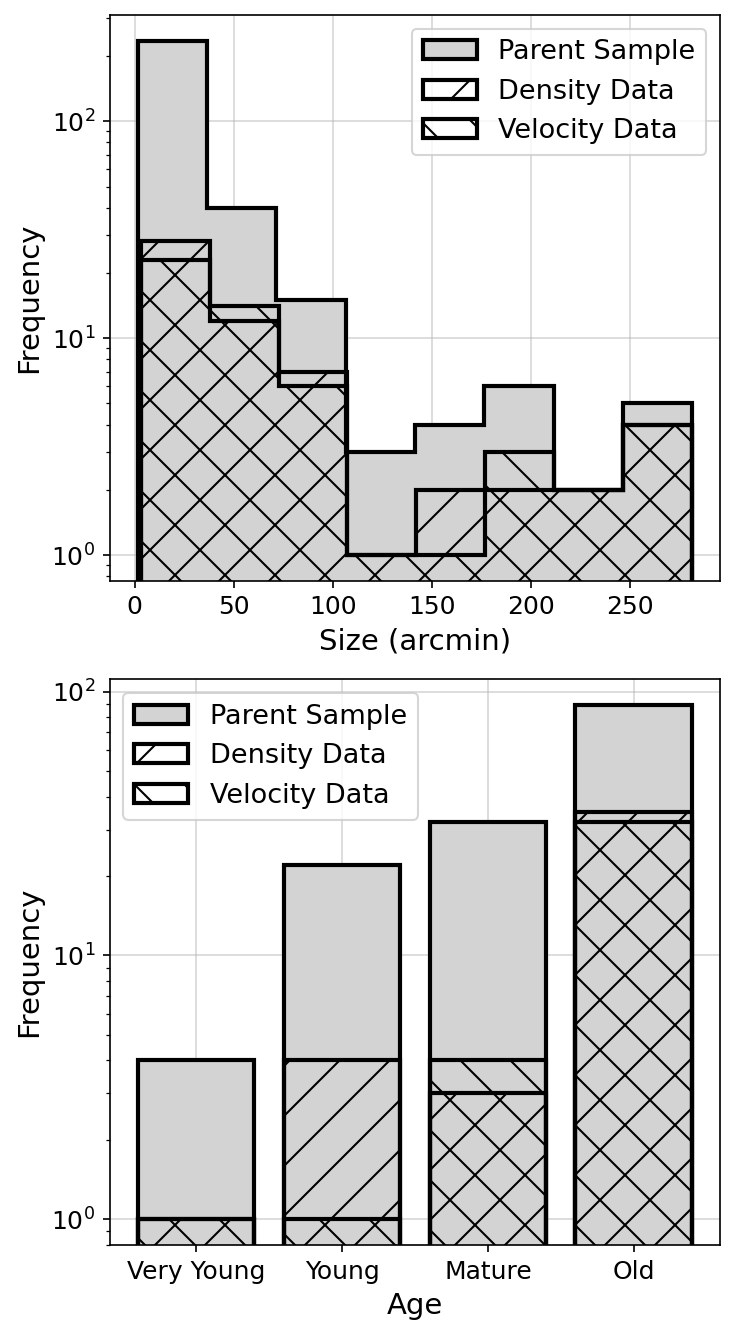}
    \caption{Distribution of Galactic SNRs with available density and velocity measurements with respect to the parent sample in the catalog of \protect\citet{Green}, in terms of their angular size and age. 'Very Young', 'Young', 'Mature', and 'Old' refer to SNRs with ages $<1$ kyr, $1-5$ kyrs, $5-10$ kyrs, and $>10$ kyrs, respectively.}
    \label{fig:bias}
    \end{figure}
    
    The identification of relevant publications was performed by a custom mining script that searched their abstracts for measurements on optical tracers of temperature, density, or shock velocity (e.g., optical emission line measurements in key diagnostic lines like [N II], [S II], [O III], H$\alpha$, H$\beta$ or their ratios), as well as keywords associated with optical emission, such as 'H-alpha', 'Balmer' and 'optical'. While this posed the risk of missing some publications with relevant information mentioned in the body of the text rather than their abstract, examination of a small number of publications not retained by our script showed that they are irrelevant to this work. In addition, although the basis for our analysis are the publications reporting optical observations from the list of \citet{Green}, in a few cases where additional publications came to our attention (e.g. in the case of Tycho SNR) these were included in our analysis for completeness.

    After applying this filtering process, we manually searched the body of the selected publications to identify any relevant information. This information includes measurements of the remnants' shock or expansion velocities, densities and temperatures, based on a variety of methods and tracers. We ensured that measurements were firstly grouped according to the region of each SNR they referred to, and then according to the tracer and method used. Age and distance information was largely obtained from the database of high-energy (X-ray and $\gamma$-ray regimes) observations of Galactic SNRs (\citeauthor{Safi-Harb}, \citeyear{Safi-Harb})\footnote[1]{Available here: \url{http://snrcat.physics.umanitoba.ca}}, which, among other information, compiles the most up-to-date information on age and distance and is regularly updated. It is worth noting that age estimates were often highly uncertain. Temperature information was also recorded, but the majority relied on assumptions and theoretical models as opposed to direct measurements. Therefore, we do not explore temperature the same way we do shock velocity and electron density in this study.

    Our final sample consists of 64 SNRs (21\% of initial sample) possessing documented data on their physical parameters derived from optical observations. These are nearly equally distributed across the northern and southern hemispheres. More specifically, 29 objects (or 45\%) are found in the northern hemisphere and 35 (55\%) in the southern hemisphere. The majority of the objects are more mature SNRs (i.e. well above a thousand years old) while five SNRs are young (up to about a thousand years old). These objects, in ascending age, are Cas A (340 yrs), Kepler's SNR (416 yrs), Tycho's SNR (451 yrs), the Crab Nebula (966 yrs) and SN 1006 (1017 yrs), which are also well known for their historical value. Table \ref{tab:table} in Appendix \ref{app:objects} shows the objects in our sample along with the number of publications from optical observations retained for each object. For the remaining 231 objects there are either no available publications or no optically-based physical parameters.

    The data extracted from these publications are presented in a consolidated table, which can be found in Appendix \ref{app:dataset}.

    \subsection{Selection bias}
    Since our analysis is based on an extensive literature survey, it can be subject to selection biases. Such biases could be related to the brightness of the objects (e.g., spectroscopic measurements are easier to obtain for brighter objects) or the selection of objects for detailed observations (observer bias; i.e., the tendency to observe the most interesting, brighter, or, in general, objects that stand out in one way or another). In order to assess this effect, we compare the distribution of objects with available measurements with respect to the parent sample (i.e., the catalog of \citeauthor{Green}, \citeyear{Green}), in terms of their angular size and age.  Although comparison in terms of physical size offers better insights into potential biases with respect to the physics state of SNRs, the lack of reliable distance measurements for a large fraction of the SNR population (25\% of the parent sample) would bias this comparison. Since many of the SNRs are asymmetric, the size is calculated as the geometric mean of the  major and minor axis reported in the catalog of Green.

    In Fig. 1 we present the distributions of the parent sample of \citet{Green} and the sample of SNRs with density/velocity measurements available in the (optical) literature, in relation to their sizes and ages. We see that the samples with density/velocity data fairly cover the parent sample apart from the small range of SNRs with sizes $\sim$100-150 arcmin. This however is not expected to bias our results significantly since it does not fall on any specific region of the parameter space. Furthermore, the very young SNRs with available density/velocity  measurements cover only a small part of the overall youngest SNRs. This is due to the fact that the latter are either dominated by a pulsar-wind nebula or in many cases they do not exhibit strong optical emission (c.f. Leonidaki et al., to be submitted).

    On the other hand, the very young SNRs ($<$1 kyr) account for $\sim$6\% of the overall parent sample. Although they seem to be biased against the overall sample, this is a fair representation since the low supernova rate and the brief duration of the very young SNR phase justify the small number of observed, very young SNRs. However, this does introduce a systematic bias due to the scarcity of remnants in early phases of evolution compared to later stages.
    
    Nonetheless, our analysis provides the first systematic census of physical parameters for optically emitting SNRs, and appears to provide a representative picture of the SNR populations regardless of angular size, and for SNRs with ages older than a few thousand years.  

%% file: sections/section_3.tex
\section{Data homogenization}
\label{sec:3}
\subsection{Data uncertainties}
\label{sec:uncertainties}
Our dataset comprised various data types, each with its unique characteristics and ways of recording uncertainties. These include upper and lower limits, value ranges and approximate values (i.e., values are provided like 100-150 or $\simeq$90), among regular values with errors. Dealing with this diverse range of data types was a significant challenge, as we could not treat all data uniformly, while simply ignoring the measurements with  unconventional uncertainties would result in a substantial loss of valuable information and a biased sample, which is particularly critical considering our already limited sample. In order to overcome this challenge, we developed a methodology that effectively incorporates all the available data, allowing us to extract as much information as possible while considering the wide range of uncertainties presented in the literature.

\begin{table}
    \caption{An indicative example of a set of different types of shock velocity measurements from a single region of an SNR.}
    \label{tab:abc}
    \centering
    \begin{tabular}{cccc} 
    \hline\hline
        A/A & Sh. Velocity (km s$^{-1}$) & Data Type & Paper \\
        \hline
        1 & $>$90 & Lower limit & A \\
        2 & $<$110 & Upper limit & B \\
        3 & 90±20 & Value with error & C \\
        4 & 90-110 & Value range & C \\
        5 & $\approx$100 & Approximate value & D \\
    \hline
    \end{tabular}
    \smallskip
\end{table}

In Table \ref{tab:abc} we present indicative shock velocity measurements for a single region of an SNR from multiple publications as well as from different methods. For the majority of objects examined in this study regions are clearly defined either by cardinal directions with reference to the limb or the center of the SNR, or by the interior vs exterior distinction, or by names recognized by the community (e.g., knot "g" in Tycho's SNR). However, in cases where multiple publications are considered, it is not always specified whether the region being referred to is the same, and for simplicity all regions within a single object are treated as unique. In fact, apart from the cases of distinct features such as knots it is more likely that the spectroscopic information  is not obtained from exactly the same regions.

Measurements for the same region may come from different methods reported in the same publication (e.g., paper C) or separate publications (e.g., papers A, B, and D).  Our goal is to obtain a summary of the different measurements while accounting for their uncertainties, even if they are recorded with different methods. Because of the different nature of the uncertainties in each measurement we cannot use traditional methods of error propagation. 

To overcome this limitation, we employed a Monte Carlo method by drawing values from appropriate probability distributions for each one of the measurements in Table \ref{tab:abc}. The distribution assumed for the Monte Carlo sampling depended on the type of uncertainty (c.f. Table \ref{tab:abc}). 

    \begin{figure*}
    \centering
    \includegraphics[width=\textwidth]{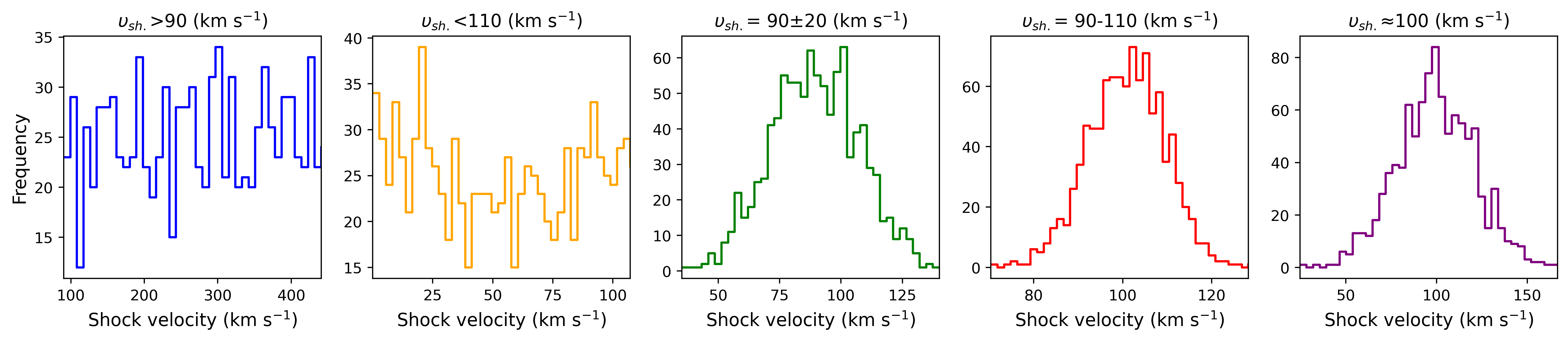}
    \includegraphics[width=\textwidth]{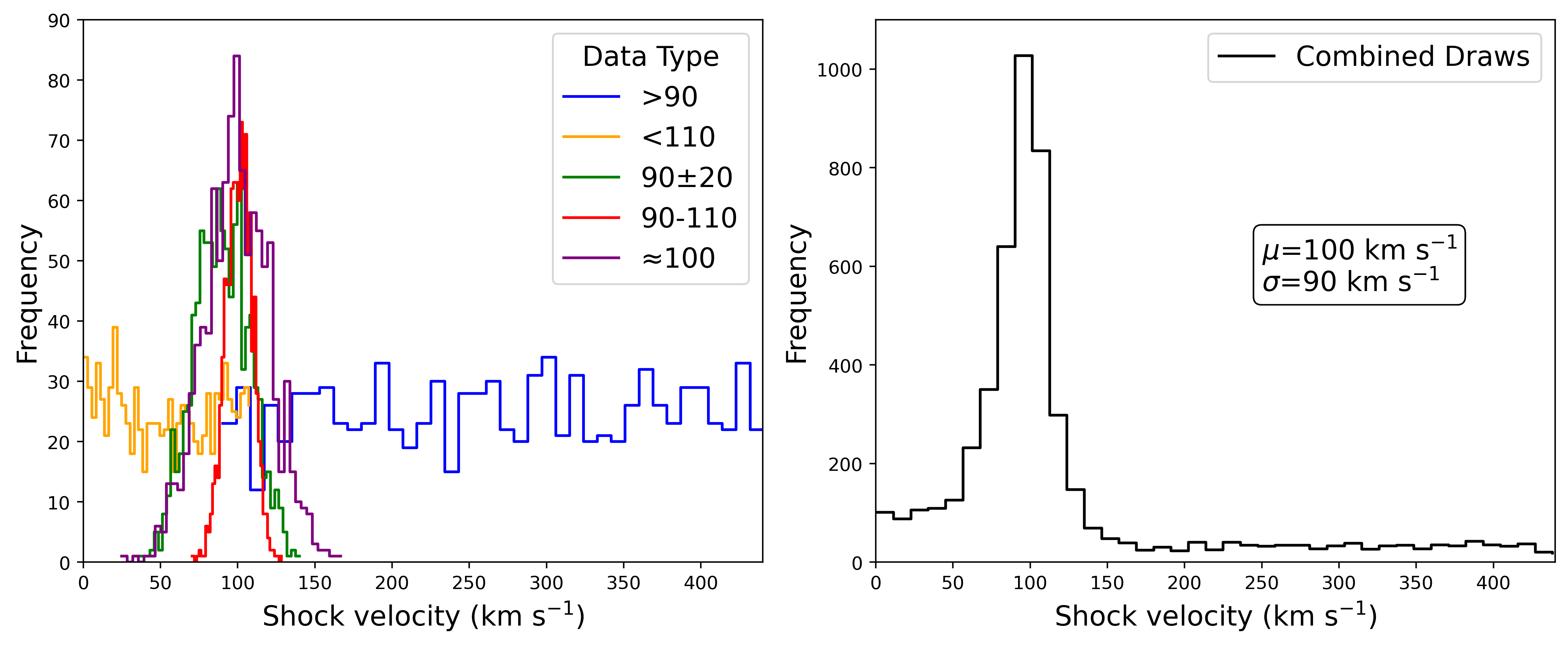}
    \caption{Visualisation of the Monte Carlo sampling process used in this work. \emph{Top}: Each histogram represents a sample of a thousand draws from appropriate distributions relative to the data type of each indicative shock velocity measurement of Table \ref{tab:abc} which is displayed at the top in units of km s$^{-1}$. \emph{Bottom:} \emph{Left}: The histograms combined into a single plot. \emph{Right}: Combining the drawn samples for each shock velocity measurement into a single histogram and calculating the median value and standard deviation of the final sample.}
    \label{fig:monte_carlo}%
    \end{figure*}

\subsubsection{Values with errors}
In the case of data with symmetric errors we assume that they follow a Gaussian distribution with $\sigma$ equal to that reported in Eq. \ref{eq:sigma}. If uncertainties in a different confidence interval are reported they are converted to the 1$\sigma$ error assuming a Gaussian distribution.

In the case of asymmetric errors we assumed two different Gaussians, one for the left error and one for the right error.

\subsubsection{Value ranges}
In the case of value ranges, we used a Gaussian distribution with a mean value equal to the midpoint of the value range, and standard deviation calculated from the following formula, assuming each time that the value range corresponds to the FWHM of the distribution:
\begin{equation}
\label{eq:sigma}
\sigma=\frac{FWHM}{2\sqrt{2\ln2}}=\frac{\text{(value range)}}{2\sqrt{2\ln2}}
\end{equation}

\subsubsection{Approximate values}
Approximate values, where uncertainties are not reported, were handled in a similar fashion, using a Gaussian distribution with mean equal to the approximate value, and assuming a 25\% error. The standard deviation is then calculated employing Eq. \ref{eq:sigma} again, assuming a FWHM equal to the value range implied by the 25\% error.

\subsubsection{Upper and lower limits}
For lower limits, we employed a Heaviside step function that extends up to five times the value of the lower limit, while, for upper limits, the Heaviside function extends down to zero. Given the lack of any information on the actual probability distribution of the parameter of interest, this assumption is the least informative while it does not overly bias the final results to extreme values. \\

We provide an example to visualise the measurement consolidation process. Table \ref{tab:abc} shows a set of hypothetical shock velocity measurements of various data types in a single region of a SNR. Fig. \ref{fig:monte_carlo} represents the visualization of the Monte Carlo sampling process. The top panel shows histograms of a thousand draws for each measurement. At the bottom, the left panel shows the histograms combined into a single plot while the right panel combines the draws into a single histogram. We calculate the mean or median value, as well as the standard deviation of the aggregated distribution to get a measurement for this region representing all available information.

\subsection{Data obtained with different methods}
While electron density is generally obtained through emission line ratios that give consistent results for a specific region, the shock velocities can be obtained from different methods that may result in systematically different estimates. For example, in the case of younger objects, where shocks are faster, Doppler shift or line broadening effects are the most direct and convenient ways to measure shock velocity. On the other hand, in older objects (i.e. the majority of our sample), where shocks are slower and therefore Doppler shift and line broadening effects are not as significant, we rely mostly on the ionization structure of the shocked gas to retrieve information on the shock velocity. Most commonly, this structure is identified through emission line ratios.

It is important to mention that different methods can be used for SNRs of different ages or in different regions within them. Since different methods for shock velocity measurements may have systematic biases, they are considered independently in the following analysis to control for these biases. This is achieved by grouping data obtained from the same remnant by method. For instance, if we have a total of 5 shock velocity measurements for an object, two of which were measured using method A and the rest with method B, the MC sampling approach described earlier is applied independently, once for the measurements derived with method A and once for those derived from method B. This means that, for each method a  separate median value and standard deviation is computed by combining the independent MC draws for each measurement with that method (depending on the type of reported uncertainty as described in the previous section).  In the end, we obtain as many estimates for each parameter per object as the number of different velocity measurement methods used in that object.

%% file: sections/results.tex
\section{Results}
\label{sec:4}
Using the data presented in Table \ref{tab:dataset} and the approach described in Section \ref{sec:3}, we combine the obtained measurements of electron density and shock velocity for each region and each object. Based on these values, we performed a statistical analysis of these physical parameters within each SNR (for those that have measurements in multiple regions) and for the overall SNR population of our Galaxy, considering the combined measurements for all regions within an object. Through these statistics, we explore the interior of SNRs, what  information it conveys about the stage of evolution, and to what extent this evolution deviates from theoretical models.

\vspace{-0.3cm}
\subsection{Intra-object statistics}
\begin{landscape}
  \begin{figure}
    \centering
    \includegraphics[width=1.35\textwidth]{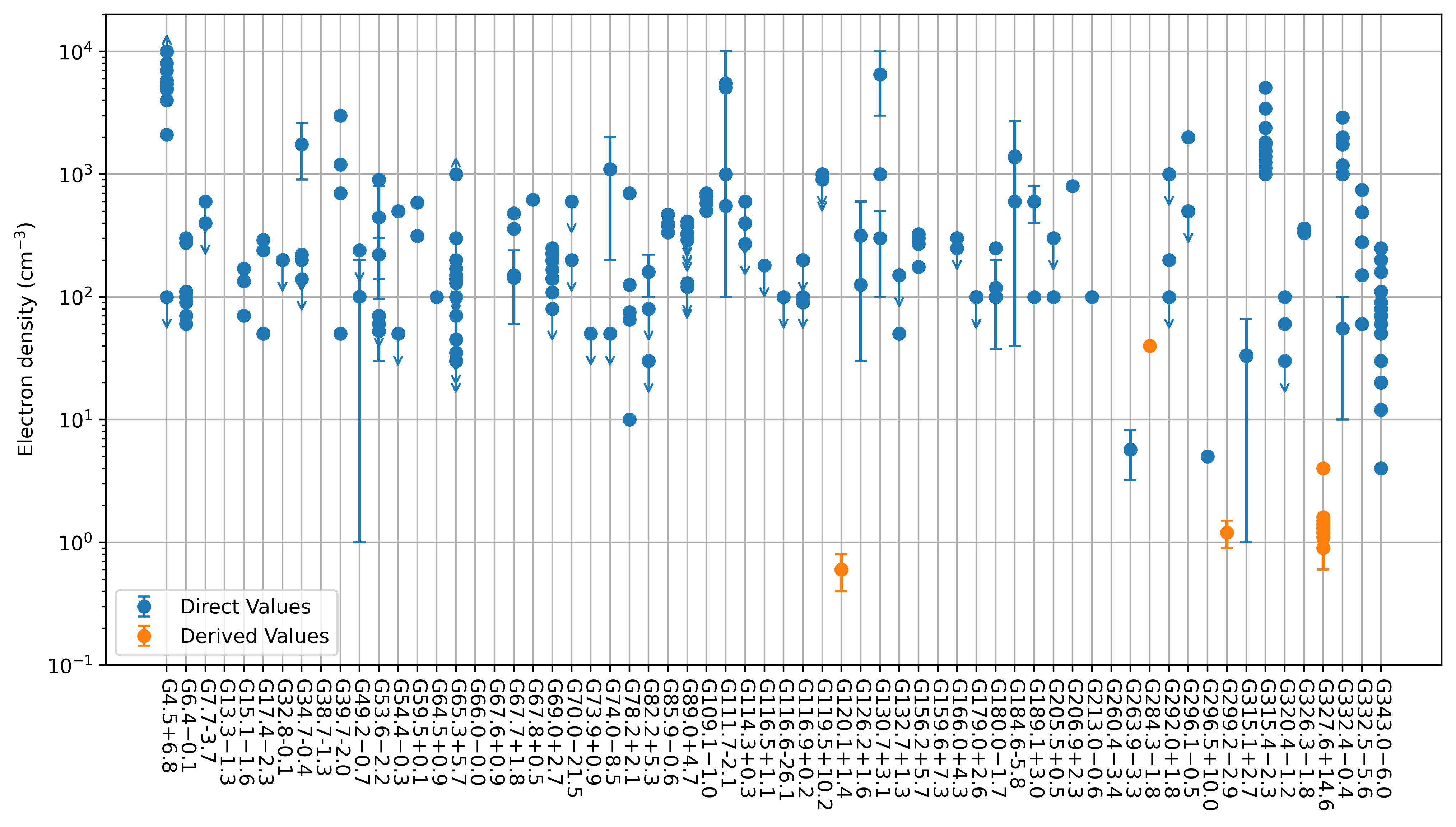}
    \caption{Electron density measurements for the different objects in our sample. Each object may have different measurements representing different regions within the remnant. Multiple measurements for a single region were grouped together according to a Monte Carlo method described in Sec. \ref{sec:3}. Objects G13.3-1.3, G38.7-1.3, G66.0-0.0, G67.6+0.9, G120.1+1.4, G159.6+7.3, G260.4+3.4, G284.3-1.8, and G327.6+14.6 do not have available electron density measurements in the literature. However, for a few objects, namely G120.1+1.4, G284.3-1.8, G299.2-2.9, and G327.6+14.6, pre-shock densities are reported and these are then multiplied by a factor of 4 to obtain an estimate for the post-shock density (see second to last paragraph of Sec. \ref{sec:4}).}
    \label{fig:den_per_obj}
  \end{figure}
\end{landscape}

\begin{landscape}
  \begin{figure}
    \centering
    \includegraphics[width=1.35\textwidth]{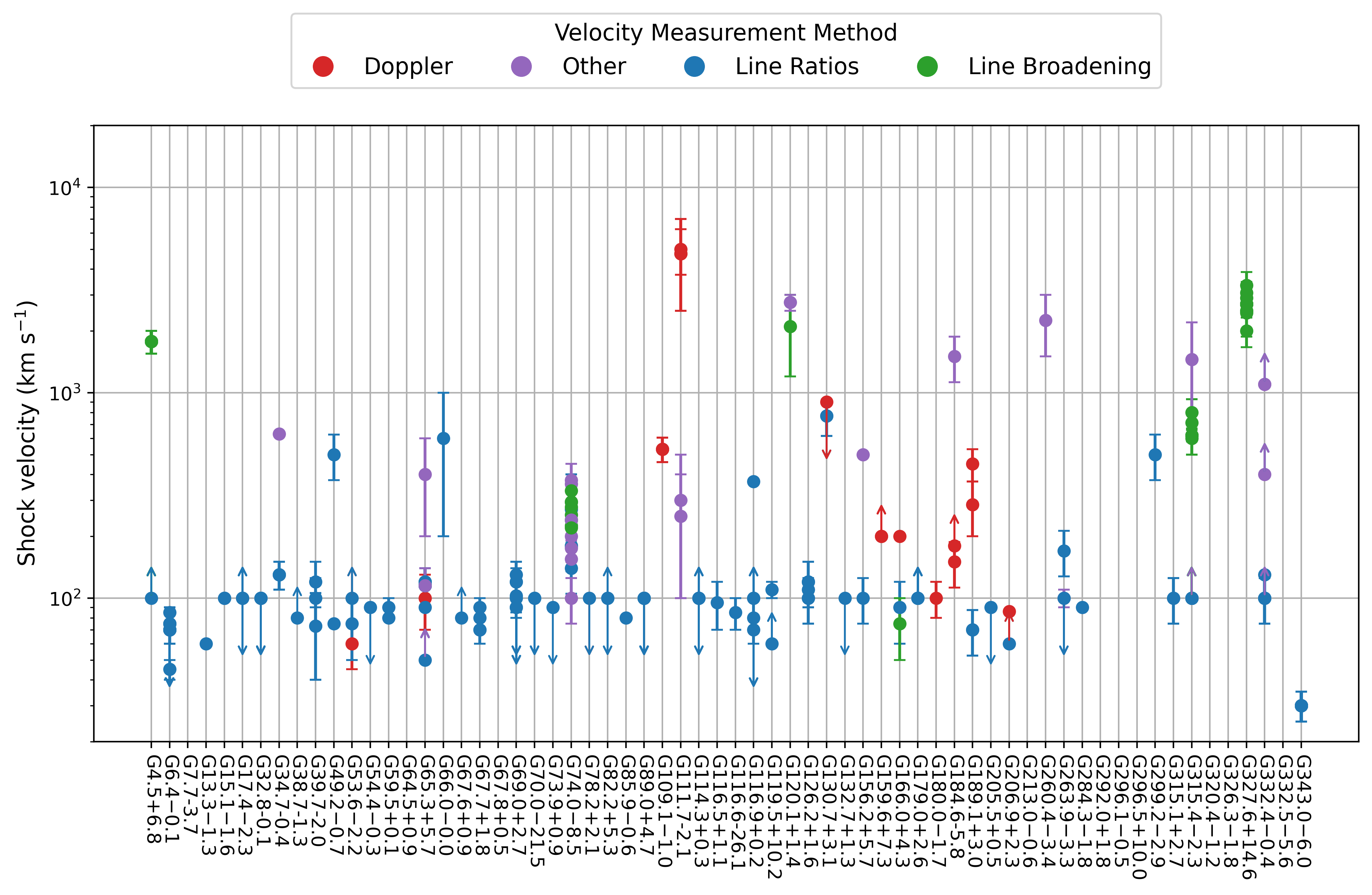}
    \caption{Shock velocity measurements for the different objects in our sample, color-coded with respect to the measurement method used. Each object may have different measurements representing different regions within the remnant. Multiple measurements for a single region were grouped together according to a Monte Carlo method described in Sec. \ref{sec:3}. Objects G7.7-3.7, G64.5+0.9, G67.8+0.5, G213.0-0.6, G292.0+1.8, G296.1-0.5, G296.5+10.0, G320.4-1.2, G326.3-1.8, and G332.5-5.6 do not have available shock velocity measurements in the literature.}
    \label{fig:vel_per_obj}
  \end{figure}
\end{landscape}

\begin{figure*}
     \vspace{-0.4cm}
     \centering
     \includegraphics[width=\textwidth]{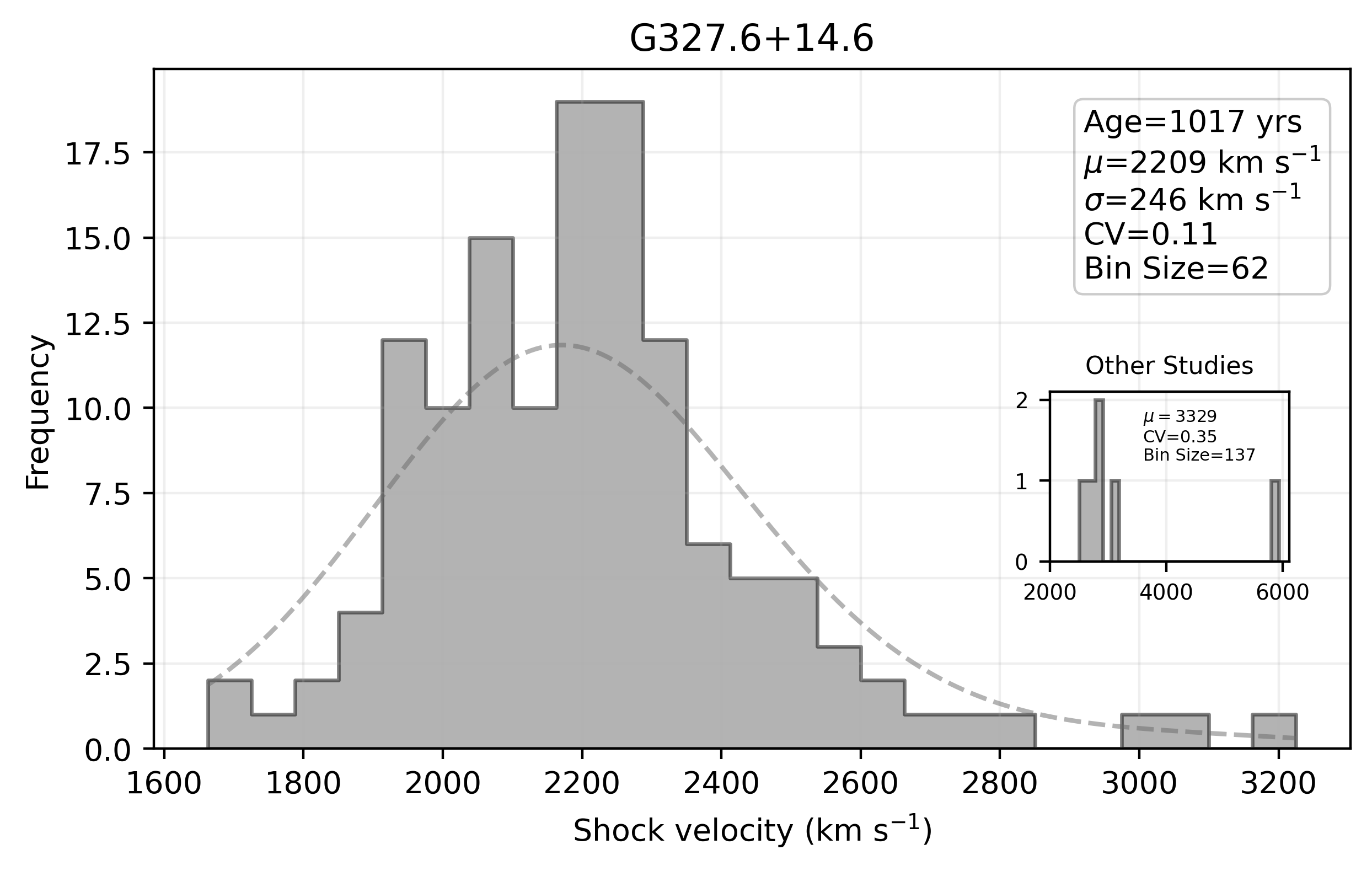}
     \includegraphics[width=\textwidth]{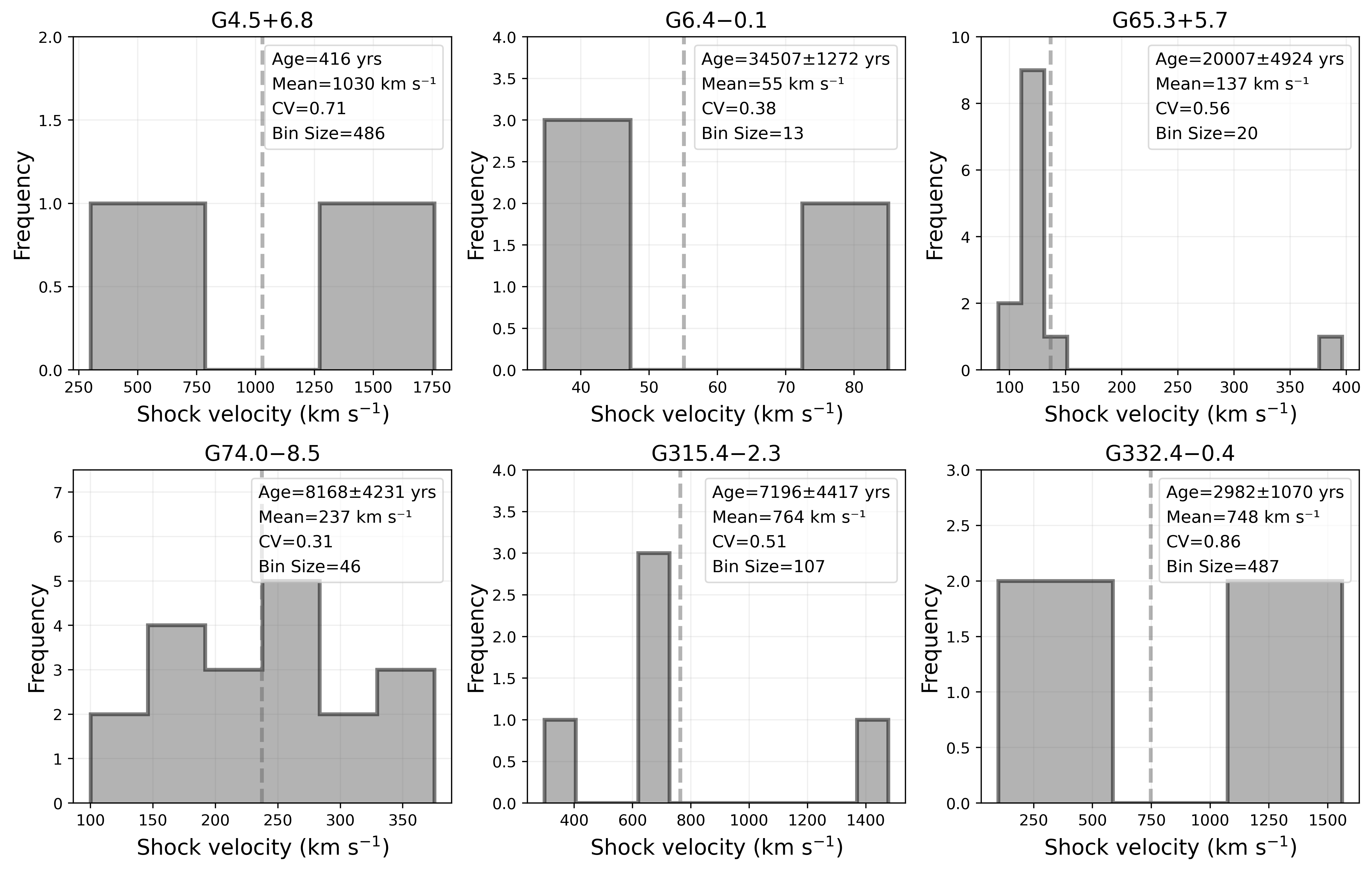}
     \caption{Shock velocity distribution within individual objects. Available measurements are grouped by region of SNR, so frequencies add up to the total number of regions observed per remnant. Legends show the age (in yrs) of each object, mean velocity (in km s$^{-1}$) and coefficient of variation (dimensionless) of the sample, as well as the bin size of each histogram in units of km s$^{-1}$. The latter has been adapted based on the dynamic range of the measurements. \emph{Top}: Shock velocity measurements for 133 distinct regions within G327.6+14.6 from \protect\cite{2013Sci...340...45N}. The inset shows measurements from independent studies.}
     \label{fig:vel_distros}%
     \end{figure*}

\begin{figure*}
     \centering
     \includegraphics[width=\textwidth]{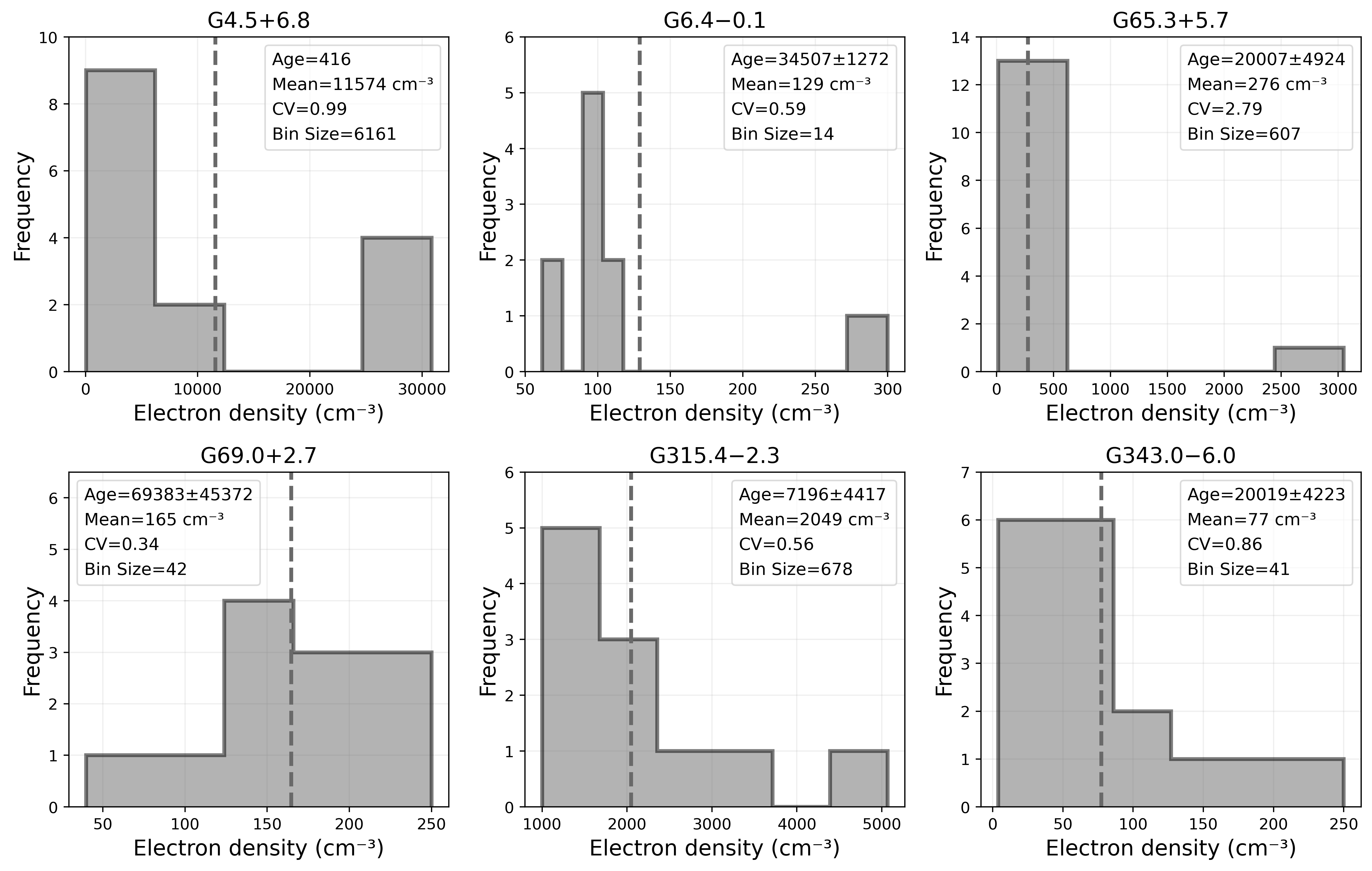}
     \caption{The electron density distribution within individual objects. The histograms were created as in Fig. \ref{fig:vel_distros} and the legends give the same information. The mean value and bin size of each histogram are given in units of cm$^{-3}$.}
     \label{fig:den_distros}%
     \end{figure*}

In Figs. \ref{fig:den_per_obj} and \ref{fig:vel_per_obj} we present the electron density and shock velocity measurements for every remnant in our sample. When measurements for multiple regions are available these are presented separately. One of our goals is to investigate, in a quantitative manner, the variation of the physical parameters within these remnants. This will allow us to obtain a picture of the physical conditions in their interior environment. For that reason, we performed a statistical analysis for SNRs of our sample with a sufficient number of measurements (>10) from different regions (see Figs. \ref{fig:vel_distros} and \ref{fig:den_distros}, as well as Table \ref{tab:stats}). To explore the variation within these remnants, we also calculated the coefficient of variation (CV) for each object, which is a standardized measure of dispersion of a distribution and is defined as the ratio of the standard deviation $\sigma$ to the mean $\mu$. Measurements from the same region were averaged, so histogram frequencies add up to the total number of distinct regions observed in each remnant. In particular, from our literature review, we found an object, SN1006 (G327+14.6), with shock velocity measurements for 133 distinct regions (see \citeauthor{2013Sci...340...45N} \citeyear{2013Sci...340...45N}). Unfortunately, few objects have been studied extensively enough, so the majority of these histograms are not well sampled. Yet, they still give a first picture of the variations in the physical param\-eters within an object, reflecting the variations in the medium surrounding SNR sites. The insights that can be drawn from these results are further discussed later.


\begin{table}
        \small
        \centering
        \caption{Mean values, standard deviations and coefficients of variation for objects with multiple measurements (>10) of shock velocity and electron density from different regions. The objects appear in ascending order of age.}
        \label{tab:stats}
        \begin{tabularx}{\columnwidth}{lXXXXXXX}
            \hline\hline
            Object \phantom{\Large I} & Age (kyrs) & \multicolumn{3}{c}{$\upsilon_{sh.}$ (km s$^{-1}$)} & \multicolumn{3}{c}{$n_e$ (cm$^{-3}$)} \\ \hline
            & & $\mu$ & $\sigma$ & $\sigma/\mu$ & $\mu$ & $\sigma$ & $\sigma/\mu$ \\ 
            G4.5+6.8    & <1        & 1034 & 735  & 0.71 & 11424 & 11147 & 0.98 \\
            G327.6+14.6 & $\simeq$1 & 3372 & 1262 & 0.37 & -     & -     & -    \\
            G315.4-2.3  & >2    & 757  & 377  & 0.50 & 2049  & 1153  & 0.56 \\
            G332.4-0.4  & >7    & 742  & 645  & 0.87 & -     & -     & -    \\
            G74.0-8.5   & >10     & 244  & 76   & 0.31 & -     & -     & -    \\
            G343.0-6.0  & $\sim$20  & -    & -    & -    & 77    & 67    & 0.86 \\
            G65.3+5.7   & >20     & 137  & 77   & 0.56 & 264   & 730   & 2.77 \\
            G6.4-0.1    & >30     & 55   & 21   & 0.38 & 129   & 76    & 0.59 \\
            G69.0+2.7   & $\le$60   & -    & -    & -    & 164   & 56    & 0.34 \\
            \hline
        \end{tabularx}
\end{table}

\vspace{-0.3cm}

\subsection{Population statistics}
\label{sec:pop_statistics}
\begin{figure}
    \centering
    \includegraphics[width=\columnwidth]{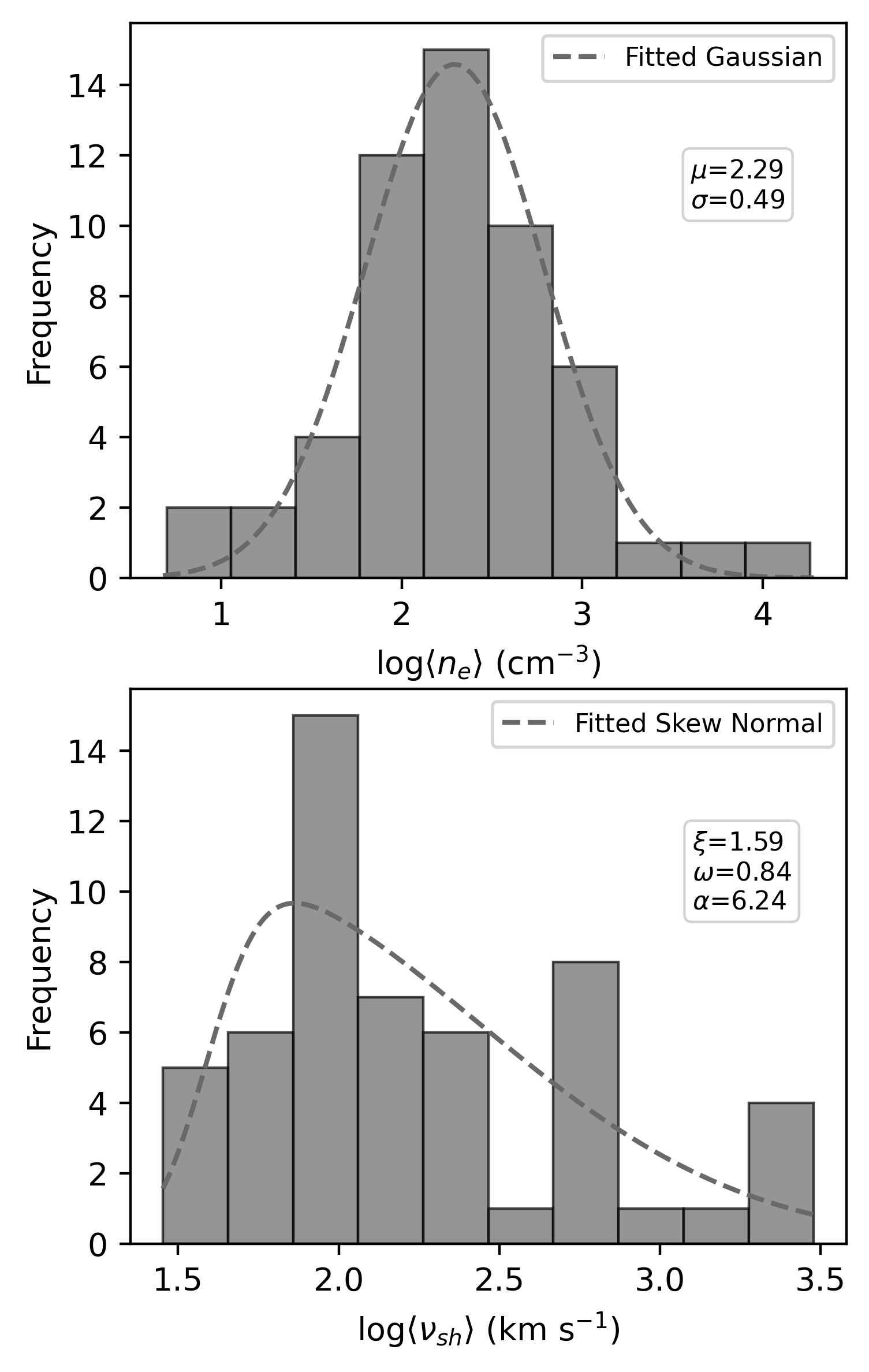}
    \caption{Physical parameter distributions for the entire sample of Galactic SNRs. \emph{Top}: The electron density distribution of the population is fitted by a log-normal distribution that has a mean $\log(n_e)=2.29$ (cm$^{-3}$) and a standard deviation of $0.49$ (cm$^{-3}$). \emph{Bottom}: The shock velocity distribution of the population is fitted by a skew log-normal distribution located at $\log(\nu_{sh.})=1.59$ (km s$^{-1}$) and having a scale of $0.84$ and a skewness of $6.24$.}
    \label{fig:pop_distros}
    \end{figure}

Our next goal is to perform statistics on the entire sample. We used the shock velocity and electron density measurements and we created histograms of each physical parameter (see Fig. \ref{fig:pop_distros}). To do this, all available measurements were grouped by object (regardless of region), following the methodology described earlier, in order to obtain combined mean values for each object. The electron density distribution is modeled with a log-normal distribution with a mean $\log(n_e)=2.29$ (cm$^{-3}$) and a standard deviation of $0.49$ (cm$^{-3}$), whereas the shock velocity distribution with a skew log-normal distribution located at $\log(v_{sh.})=1.59$ (km s$^{-1}$) with a scale factor of $\omega=0.84$ and a skewness of $\alpha=6.24$. Skewness is a measure of the asymmetry of the distribution about its mean, the location parameter determines the "location" or shift of the distribution and the scale parameter determines its spread. More specifically, the probability density function (PDF) of a skew normal distribution with a skewness parameter $\alpha$ is given by 
\begin{equation}
\label{eq:10}
f(x) = 2\phi(x)\Phi\left(\alpha x\right),
\end{equation}
where $\phi(x)=\frac{1}{\sqrt{2\pi}}e^{-x^2/2}$ is the standard normal PDF and $\Phi(x)=\int_{-\infty}^x\phi(t)dt=\frac{1}{2}\left[1+erf\left(\frac{x}{\sqrt{2}}\right)\right]$ is the cumulative distribution function (CDF), where "erf" is the error function. To add location and scale parameters, one makes the usual transform $x\rightarrow\frac{x-\xi}{\omega}$ and divides the PDF by $\omega$, where $\xi$ and $\omega$ are the location and scale parameters, respectively. When $\alpha=0$, $\xi=0$ and $\omega=1$ the distribution is identical to a normal distribution centered at zero. Since the PDF is normalized, the curves shown in Fig. \ref{fig:pop_distros} are rescaled based on the total number of objects and the width of each histogram bin. 

Next, we explore SNR evolution, i.e. the relationship of the shock velocity with age. We first homogenized our data according to the approach described in Section \ref{sec:3}, where we sampled values from probability distributions corresponding to each uncertain measurement (i.e. upper/lower limit, value range, etc.). The drawn samples were grouped by method of shock velocity measurement. This is done to minimize biases associated with certain methods that tend to have systematic offsets. Finally, we took the median value and standard deviation of each group of samples and plotted the results color-coded with respect to the logarithm of the average post-shock electron density of the remnants. The final number of points on the plot corresponds to the number of objects with at least one velocity and age measurement times the number of methods per object (c.f. Fig. \ref{fig:cioffi_curves}).

\subsection{SNR evolution}
SNRs are dynamical objects, i.e. they expand and interact with their surrounding medium while they evolve. Their evolution has been modeled extensively in different levels of complexity (see e.g., \citealt{1999ApJS..120..299T}; \citealt{Bamba_2022}, for reviews). However, these models generally focus on individual evolutionary phases. In our analysis we consider the evolutionary models of \cite{Cioffi}, which take into account all evolutionary stages throughout their life. We used this theoretical model and compared it with our observational data in order to study the relation between shock velocity and age, given specific densities.  We focus on the adiabatic and radiative phases of the model since the optically emitting SNRs we are interested in are in either of the two phases.

According to \cite{Cioffi}, the transition from the Sedov-Taylor to the radiative phase takes place near the shell-formation time $t_{sf}$, when the first parcel of shocked material cools completely. As a result, a thin shell forms which "snowplows" through the ISM, driven by the pressure of the hot, roughly isobaric interior in addition to the shell's momentum (\citeauthor{Cox1972} \citeyear{Cox1972}; \citeauthor{Chevalier1974} \citeyear{Chevalier1974}). In this case
\begin{equation}
\label{eq:1}
t_{sf} = 3.61\times10^4\frac{E_{51}^{3/14}}{\zeta_m^{5/14}n^{4/7}}\,yr,
\end{equation}
where $E_{51} = E_0/10^{51}\,erg$ is the total energy of the explosion in units of $10^{51}$ erg, $n$ is the pre-shock density in units of $cm^{-3}$, and $\zeta_m$ is a factor depending on the metallicity, which is 1 for solar abundances. Hence, the transition time is given by the relation:
\begin{equation}
\label{eq:2}
t_{tr} = \frac{t_{sf}}{e}\,yr
\end{equation}
where $e$ is the base of the natural logarithm.

Based on the model of \citeauthor{Cioffi}, the shock velocity in each of these two phases as a function of the pre-shock density $n$ and the SNR age $t$ are given by:
\begin{equation}
\label{eq:3}
\nu_{ST} = 0.4(\xi10^{51}E_{51})^{1/5}\rho^{-1/5}(3.3\times10^7t(yrs))^{-3/5}\,km\,s^{-1}
\end{equation} 
for the adiabatic (Sedov-Taylor) phase, and: 
\begin{equation}
\label{eq:4}
\nu_{rad}(n,t) = \nu_{tr}(n)(\frac{4t}{3t_{tr}(n)}-\frac{1}{3})^{-7/10}\,km\,s^{-1}
\end{equation}
for the radiative phase, where
\begin{equation}
\label{eq:5}
\nu_{tr}(n) = 413n^{1/7}\zeta_m^{3/14}E_{51}^{1/14}\,km\,s^{-1}
\end{equation}
is the velocity at the beginning of the radiative phase. In the above relations, $\xi = 2.026$ is a numerical constant, $\zeta_m$ is the metallicity factor as mentioned earlier and $\rho = \mu_H n m_H = 2.3\times10^{-24}n$ gr cm$^{-3}$ is the total mass density of the ambient gas (assuming pure atomic hydrogen gas with mean mass per hydrogen atom  $\mu_H$=1, and $m_H$ the hydrogen-atom mass.

Our aim is to compare the shock-velocity data as a function of the SNR age with the theoretical curves given in the previous relations. In this comparison we use the combined data for each object as described in Section \ref{sec:pop_statistics}.  In order to minimize biases associated with different shock velocity methods when multiple methods are available we consider them independently. 

The theoretical curves for the SNR shock-velocity evolution are evaluated for three different ambient densities (10, 100, and 1000 cm$^{-3}$) covering the main body of the density measurements of the SNRs in our sample (Fig. \ref{fig:pop_distros}). Since the theoretical shock velocity also depends on the evolutionary phase we employed the following approach: We create an age grid based on the ages of the SNRs. For each of these ages and each of the considered densities we evaluate whether the object is in the adiabatic or radiative phase by comparing the age with the transition timescale $t_{tr}$ (Eq. \ref{eq:2}). If $t_{tr}$ is larger than the age then we adopt Eq. \ref{eq:3} for the shock velocity (adiabatic phase), while if it is smaller we adopt Eq. \ref{eq:4} (radiative phase).

\begin{figure*}
    \centering
    \includegraphics[width=\textwidth]{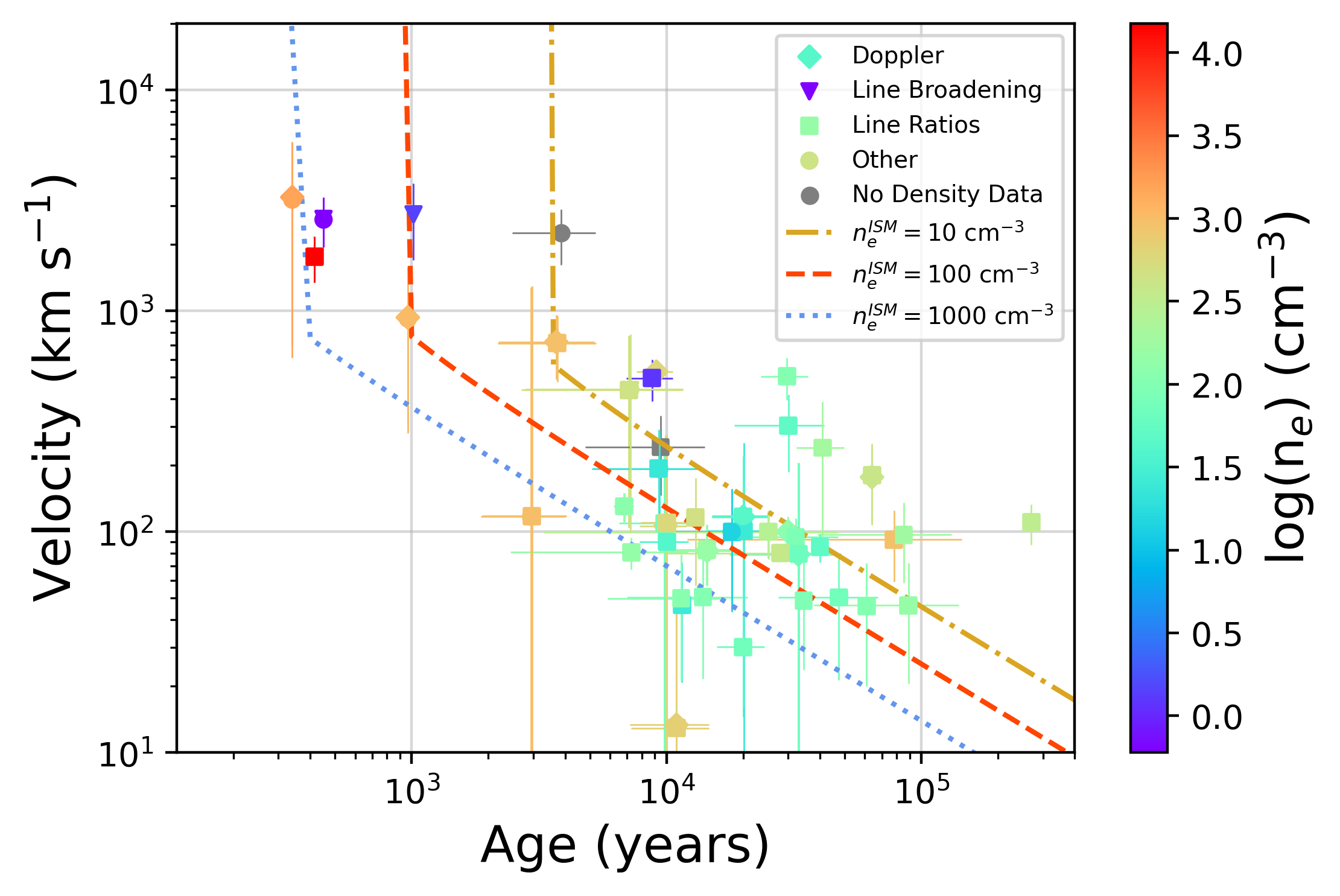}
    \caption{Shock velocity against age and theoretical lines based on the model of \protect\cite{Cioffi} for different ISM densities. Data are grouped by velocity measurement method, using the median value of the grouped samples, and color coded with respect to the average post-shock density of the groups. Errors are estimated according to the MC sampling approach described in Section \protect\ref{sec:uncertainties}.}
    \label{fig:cioffi_curves}
    \end{figure*}

Our results are shown in Fig. \ref{fig:cioffi_curves}. Each point on the plot represents a shock velocity estimate obtained with the same method for each SNR. The lines represent the model predictions for the three different ambient densities. More specifically, the nearly vertical parts represent the transition from the adiabatic (Sedov-Taylor) to the radiative phase of evolution. The data are in good agreement with the anti-correlation suggested by the curves, judging from the lack of older SNRs with large velocities (upper right locus of the plot) and objects in the lower left part of the plot. The presence of younger objects in our SNR sample is decisive in the existence of the observed trend. This will be further discussed later.

It should be emphasized that the model uses the \emph{pre-shock} instead of the post-shock density. In cases where the pre-shock density was not reported in the publications we examined, we assume that it is one-fourth of the post-shock density (i.e. the electron density we measure) following the Rankine-Hugoniot shock jump conditions (\citeauthor{Draine}, \citeyear{Draine}). Similarly, if only pre-shock density measurements are reported in the literature, as is the case for a few objects, we multiply by a factor of four to obtain an estimate of the post-shock density. These objects are represented by orange colored data points in Fig. \ref{fig:den_per_obj}. The selection of this relation is relatively inconsequential, as equations \ref{eq:1}, \ref{eq:3}, \ref{eq:4} and \ref{eq:5} governing the shock velocity depend on the pre-shock density with small exponents. Thus, variations in the choice of the relation have a limited impact.

\begin{figure*}
    \centering
    \includegraphics[scale=0.7]{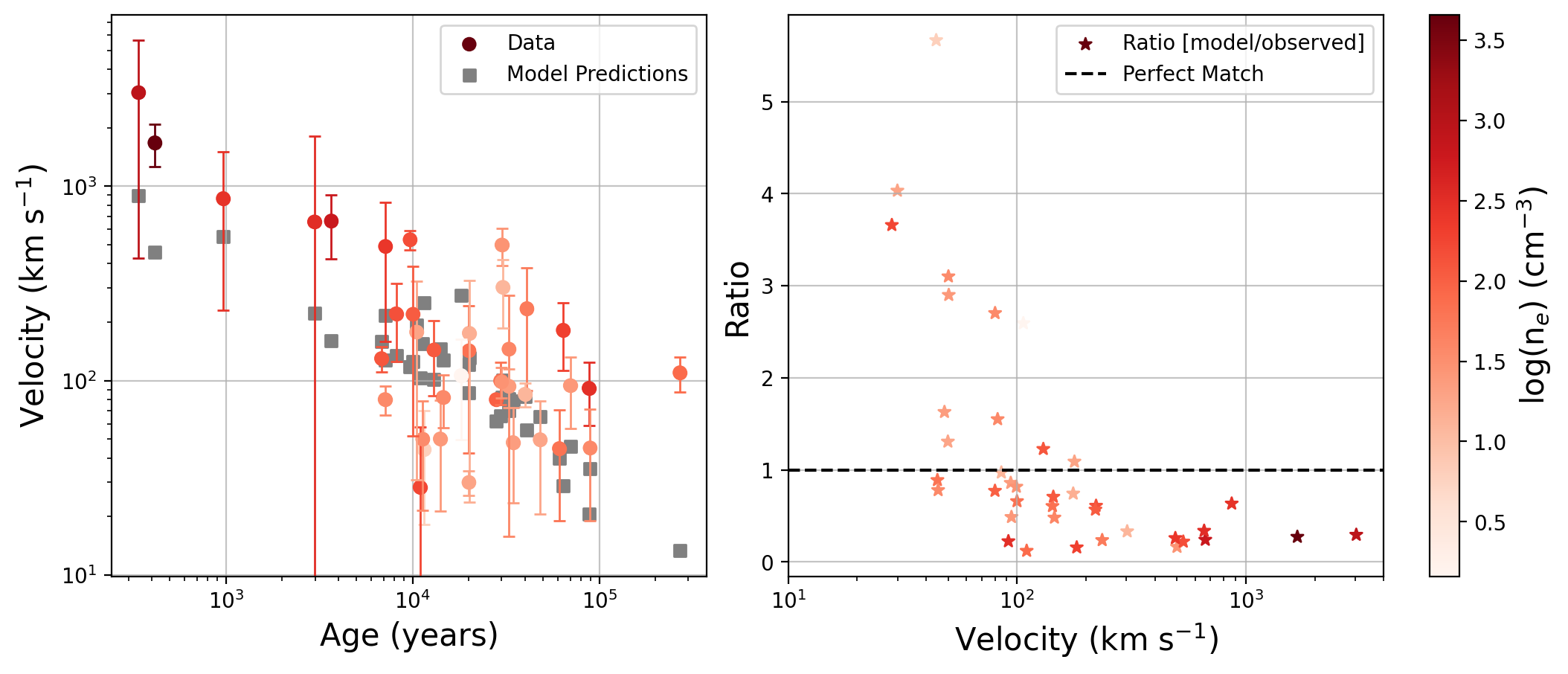}
    \caption{\emph{Left}: One-by-one comparison between shock velocity data from SNRs of a known age and the model predictions. \emph{Right}: The ratio of the model predictions to the observational data. The data are color-coded with respect to the logarithm of the ambient density of each SNR. The x-axis shows the data velocity for each point.}
    \label{fig:comparison}%
    \end{figure*}

In Fig. \ref{fig:comparison} we show a one-by-one comparison between the data and the model predictions. The ambient density, which is calculated as mentioned above, is to calculate the theoretical value of the shock velocity for each object (gray points) and also for the color-coding of the observational data. We observe partial agreement between data and model. While the observational data and the model show the same trend, there is significant scatter, and the model tends to underestimate the velocity for the youngest (and generally denser) objects, while it overestimates it for some of the lowest density objects. This results in the ratio of the model prediction to the observational data being well beyond unity for more than one-third of the shown sample (c.f. right panel in Fig. \ref{fig:comparison}). Nonetheless, the difference is at most a factor of $\sim2$ in the majority of the cases.

%% file: sections/discussion.tex
\section{Discussion}
\label{sec:discussion}
In the previous sections we presented a systematic meta-analysis of the available measurements of physical parameters such as shock velocity, electron density, and age, for SNRs in our Galaxy. This analysis considered measurements for different regions within individual objects as well as values characterizing the overall objects. We then use these results in order to explore the evolution of these parameters with SNR age, and in particular the evolution of the shock velocity.

\subsection{Distribution of shock velocity and electron density}
In Figs. \ref{fig:den_per_obj} and \ref{fig:vel_per_obj} we plot the raw electron density and shock velocity data, respectively, for each object in our sample. These data represent measurements from individual regions within the supernova remnant, with some objects having data for only one region and others encompassing several distinct regions. 

As can be seen in Fig. \ref{fig:den_per_obj}, the vast majority of sources present density values between 100 and 1000 cm$^{-3}$. This (large) scatter indicates a diversity in density between different regions within a SNR, sign of a non-uniform ISM. A few SNRs present high electron densities ($\ge$ 1000 cm$^{-3}$, G4.5+6.8: Kepler; G34.7-0.4: W44; G39.7-2.0: W50; G111.7-2.1: CasA; G130.7+3.1: 3C58, G315.4-2.3: RCW86; G332.4-0.4: RCW103), designating dense clumps and knots across their ambient medium. These density gradients can originate either from their progenitor that strongly modifies the circumstellar medium (CSM) with intense mass loss/wind-blown bubbles (e.g., Cas A; \citealt{2006ApJ...640..891Y}; RCW 86 \citealt{2011ApJ...741...96W}), or from the interaction of the SNR with the ISM (e.g., interaction with molecular cloud(s) at W44; \citealt{1994ApJ...430..757R}, jets at W50; \citealt{Bowler_2018}). The method used for electron density measurements is sulfur emission line ratios. 

The data in Fig. \ref{fig:vel_per_obj} are color-coded with respect to the method used to measure the shock velocity, including Doppler shift, line broadening and emission line ratios, and a few other less common methods. Shock velocities measured via Doppler shift and line broadening methods tend to be associated with higher values, whereas those measured via emission-line ratios correspond to lower values. The first two methods are usually preferred in young (X-ray) emitting objects, as well as regions in the shock front, where the shocked gas is expected to be heated and expanding rapidly and, thus, broadening effects and Doppler shift are expected to be significant. For example, nearly half of the high-velocity SNRs we see in Fig. \ref{fig:vel_per_obj} are young, Balmer-dominated SNRs (Kepler, Tycho, RCW 86 and SN1006), where the H$\alpha$ broad-line emission at the shock front advocates the usage of the line broadening method. On the other hand, for density ranges typical of a nebula ($\sim$100-1000 cm$^{-3}$), a different case holds. For these density ranges, emission line ratios like [OIII], are independent of density (\citealt{2008ApJS..178...20A}) but sensitive on temperature and, therefore, shock velocity diagnostics for older, cooler objects that have expanded into the ISM or inner regions of the remnant that have cooled ($\sim$ 10$^4$K). 

Consequently, for older objects (i.e. the majority of our sample) where the shock front has slowed down and the interior gas has cooled down, emission-line ratios are preferred and therefore consistent with lower values of shock velocity. On the other hand, for younger objects where the optical emission does not reflect the bulk of the hotter material, kinematic methods are more representative. X-ray line measurements would also be useful but these are strongly affected by the reverse shock. 

We notice that most of the objects have velocities around 100 km s$^{-1}$ while there are a few exhibiting shock velocities >1000 km s$^{-1}$ (i.e. Kepler, Cas\,A, Tycho, Crab, Puppis\,A, RCW\,86, RCW\,103 and SN1006), half of them being historical SNRs and all of them being relatively young ($\le$2000 yrs), as expected given their velocities. 

One would expect SNRs with higher shock velocities to present lower electron densities and vice versa (given that $\nu\sim n^{-1/2}$). Based on Figs. \ref{fig:den_per_obj} and \ref{fig:vel_per_obj}, we see that the vast majority of SNRs with velocities $\sim$100 km s$^{-1}$ correspond to a wide range of densities (100-1000 cm$^{-3}$), many measurements, and various regions. Without further investigation on individual SNRs and based solely on these plots, a qualitative comparison would lead to erroneous assumptions. However, it is interesting to further investigate the few objects that present high shock velocities ($\ge$1000 km s$^{-1}$). Four of them (i.e. Kepler, Cas\,A, RCW 86 and RCW 103) present also high densities ($\ge$1000 cm$^{-3}$). Apart from RCW 86 and RCW103, for which their high density measurements do not correspond to the same high shock velocity regions (see Appendix \ref{app:dataset}), Kepler and Cas\,A are the youngest SNRs within our sample. Even 3C58 (G130.7+3.1), the fourth youngest SNR in our sample, presents an elevated shock velocity of $\sim$850 km s$^{-1}$ while it is embedded within a dense region. A special case is the very young Tycho SNR ($\sim$450 yrs old) where its shock wave has no dense medium to confront since it expands in a low-density cavity. It appears that the energy of very young SNRs cannot be suppressed, even if they evolve within a dense medium. When we go to older SNRs, such as W44 or W50 for example, with ages $\sim$6500-7500 yrs and tens of thousands of years respectively, we see that their velocities gradually weaken when encountering dense material (W44: $\sim$650 km s$^{-1}$ and W50: $\sim$100 km s$^{-1}$). There are no other young SNRs with both high velocity and density measurements available and corresponding to the same region in order to further investigate this behavior. 

\begin{figure}
    \centering
    \includegraphics[width=\columnwidth]{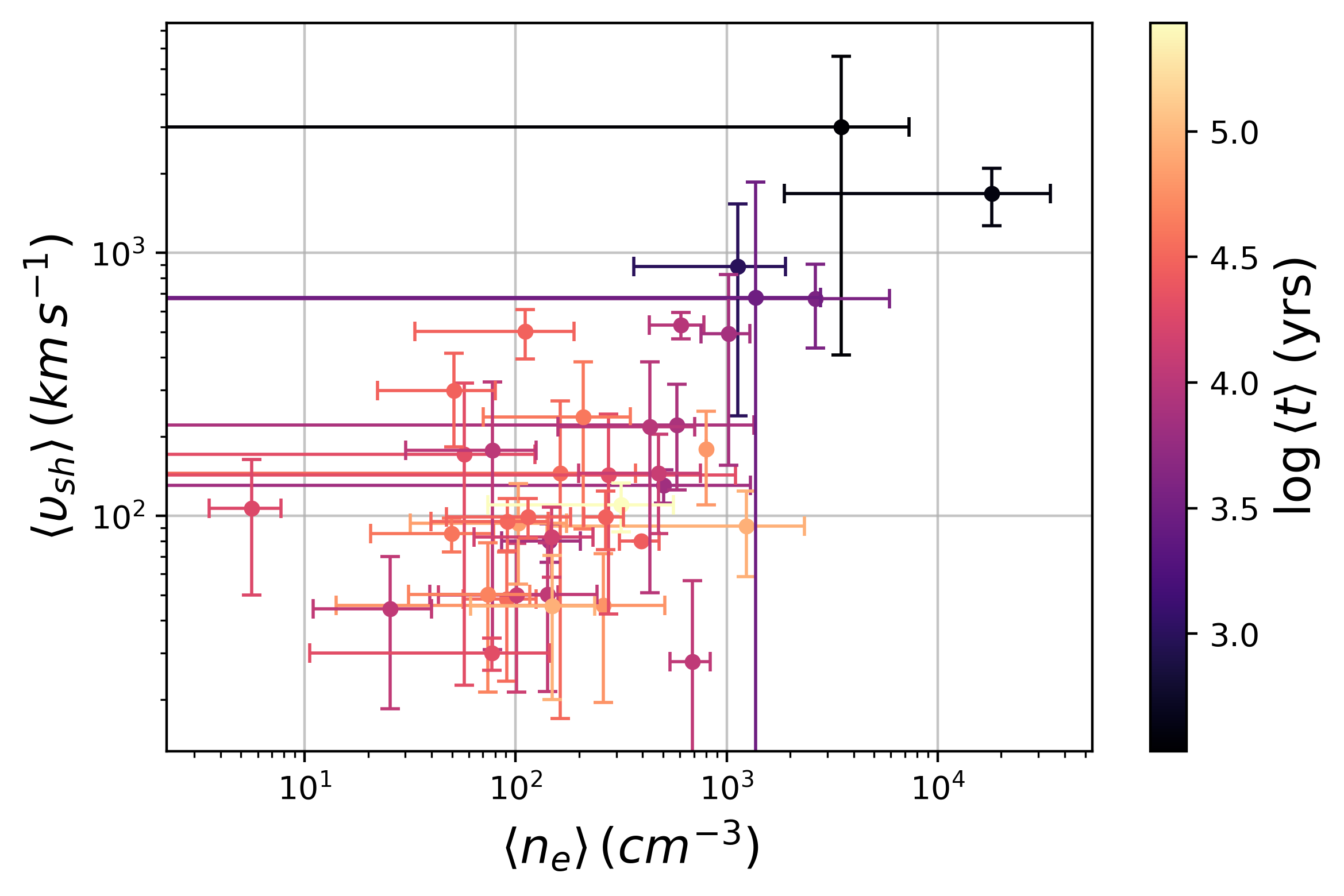}
    \caption{Mean shock velocity as a function of mean electron density, color-coded according to the logarithm of the age. A weak trend is observed, where denser objects are associated with faster shocks.}
    \label{fig:vel_vs_den}
    \end{figure}

Therefore, shock velocities of very young SNRs seem to be driven by their age while for older SNRs, velocities are more affected by their surrounding density. The relation between velocity and density occuring at waves with small width ($\nu\sim n^{-1/2}$) cannot be followed in the case of shock waves where the situation is much more complex and violent. A better visualization of this trend can be seen in Figs. \ref{fig:cioffi_curves} and \ref{fig:vel_vs_den}.

\begin{figure}
    \centering
    \includegraphics[width=\columnwidth]{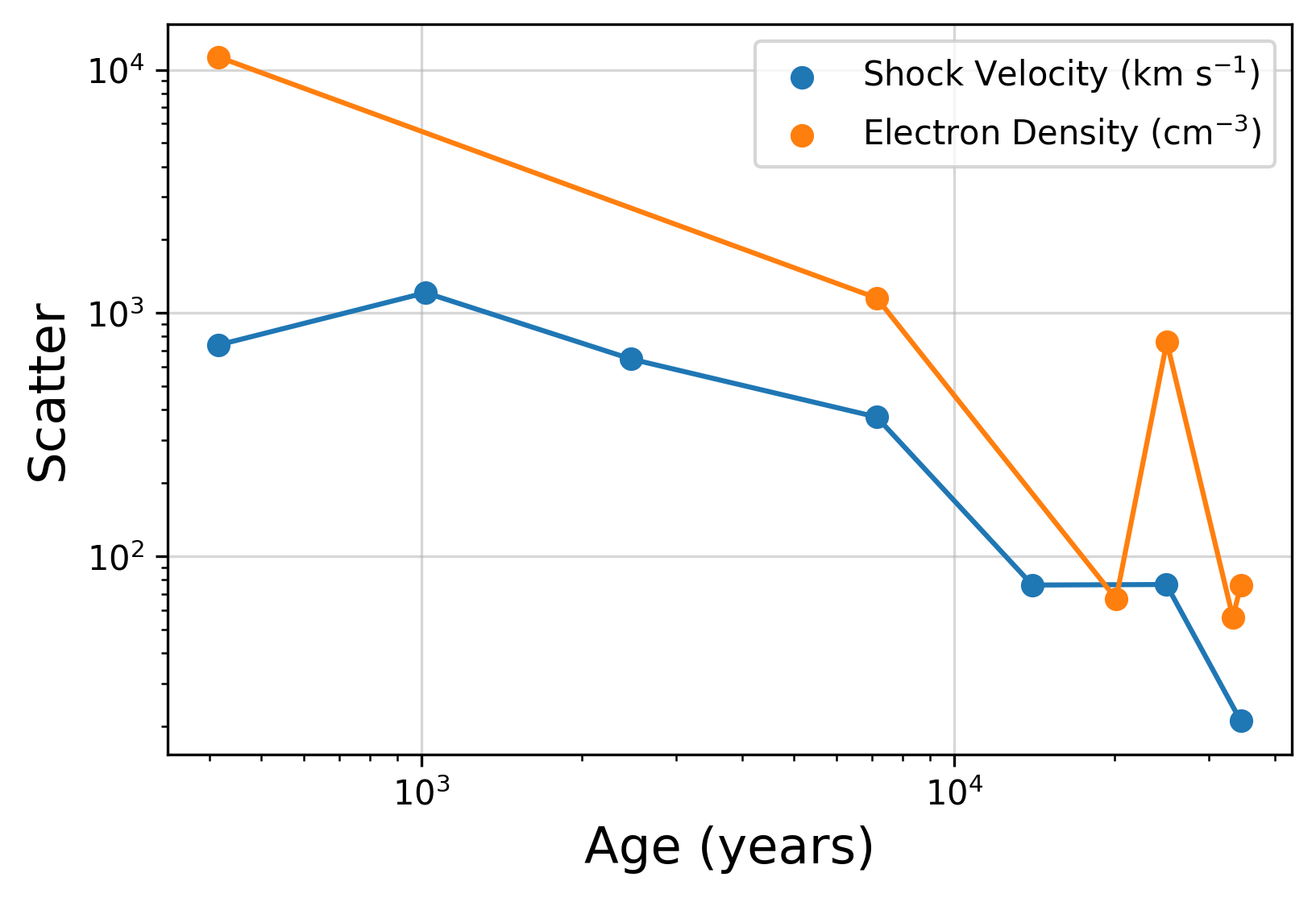}
    \caption{Scatter of intra-object shock velocity and electron density measurements with respect to SNR age. The data are based on Table \ref{tab:stats} and correspond to select objects (hence their scarcity) with less uncertain ages compared to the rest of the sample, as well as measurements across multiple regions, so that standard deviations could be calculated as a measure of scatter. Multiple measurements for a single region were handled using our sampling approach described in Section \ref{sec:uncertainties}.}
    \label{fig:scatter}
    \end{figure}

In Figs. \ref{fig:vel_distros} and \ref{fig:den_distros} we plot the distributions of shock velocity and electron density, respectively, within individual objects across different regions. From the 64 objects in our sample, we only present cases with multiple measurements (>10), estimating also the mean value and spread of these parameters. Fig. \ref{fig:scatter} shows their scatter within select remnants, whose ages are known with reasonable certainty, as a function of their age. 

The distributions in Figs. \ref{fig:vel_distros} and \ref{fig:den_distros} show significant spread, especially in younger objects, as can be inferred also from Fig. \ref{fig:scatter}. Based on the values of standard deviation, only a few remnants show relatively self-consistent shock velocity (e.g., G6.4-0.1, G65.3+5.7 and G74.0-8.5) and electron density (e.g., G6.4-0.1) values between their different regions compared to the rest of the objects (c.f. Table \ref{tab:stats}). Two of these, namely G6.4-0.1 and G74.0-8.5, appear to be interacting with a molecular cloud, while G74.0-8.5 also contains several compact sources. No interaction with a molecular cloud is reported for G65.3+5.7. All three objects are medium-old (i.e. between 10000 yrs and 36000 yrs). 

The reason why measurements of the shock velocity in different regions of a single object are not particularly self-consistent is because departures from uniformity in the structure of the ambient medium cause the shock to break up, i.e. the shock propagates with different velocities through a diverse medium, depending on its density (e.g., knotty structures, clumps, cavities), decelerating faster in denser regions compared to less dense ones.

We find that younger objects show significantly larger spread in their physical parameters compared to their older counterparts (c.f. Table \ref{tab:stats}). The gradual dissipation of the shock combined with the mixing of the initially clumpy CSM with the more uniform ISM implies that the remnant becomes more uniform as it fades out; inhomogeneities encountered in the earlier phases are smoothed out by the passing shock wave, gradually leading up to the dissipation phase of SNR evolution.

In Fig. \ref{fig:pop_distros} we plot the distributions of the physical parameters (electron density and shock velocity) of the entire sample of SNRs, as explained in Section \ref{sec:pop_statistics}. We model the electron density distribution with a log-normal distribution, as can be seen in the top panel. We expect a lower limit to the density set by the interstellar medium and an upper limit set by the mass of the progenitor star and, therefore, the amount of stellar content propelled outward into the surrounding medium. The lowest density is $\sim10$ cm$^{-3}$, consistent with typical ISM densities. However, the highest density is generally driven by the clumpiness of the CSM. According to Fig. \ref{fig:pop_distros} the middle point of the distribution falls at $log(n_e)$=2.29$\pm$0.49 (cm$^{-3}$).

When it comes to the shock velocity distribution of the sample, we model the distribution with a positively skewed log-normal function. The dampening of the distribution for greater shock velocities is expected, as it is unlikely for multiple stellar systems to have recently undergone supernova explosions simultaneously. In fact, we expect a significantly slowed down shock for most objects since most of them are more than $\sim10^4$ years old.

Finally, in Fig. \ref{fig:vel_vs_den} we observe a weak positive correlation for the mean shock velocity with respect to the mean electron density. We would expect an inversely proportional relation between the two parameters, as dense environments contribute to the deceleration of the shock. We propose that the reason behind this trend is that objects in the upper right corner are significantly younger than the rest of the sample, as can be seen from the color bar, and, therefore, have not had enough time to decelerate. This indicates that age is the driving factor of shock velocity for young, optically emitting remnants, while density is a secondary factor. This interpretation agrees with the model of \cite{Cioffi} for which the shock velocity of radiative SNRs is $\nu_{rad}\propto n^{-\frac{2}{35}}t^{-\frac{7}{10}}$. While the model does not take into account the complex structure of the CSM and ISM, it appears to describe the dependence on age rather well, particularly for younger objects. On the other hand, density appears to play a secondary role in SNR evolution, mainly in increasing the scatter. These inferences are largely reflected in more evolved SNRs having markedly slower shocks as discussed earlier, as well as increased scatter (c.f. lower right part of Fig. \ref{fig:cioffi_curves}), reflecting their interaction with and expansion into a variety of ambient media. This latter point is discussed in detail in the following section.

\subsection{Evolution of physical parameters }
We find that there is an anticorrelation of the SNR shock velocity with age, which is in good agreement with expectations from theoretical models. Our sample covers a wide range of ages ($\sim300$\,yr to $\sim10^{5}$\,yr). The presence of the youngest SNRs (e.g., Cas A, Kepler, Tycho, Crab Nebula and SN 1006) plays a decisive role in this study since they accentuate the trend and probe the diverse environment in the immediate vicinity of the explosion. A trend is also present for older objects but with larger scatter. This is because remnants, especially young ones, expand into an ambient medium whose density and structure varies significantly from remnant to remnant and from region to region within the same remnant (e.g., \citealt{2012A&A...537A.139C}; \citealt{2013MNRAS.435.1659C}; \citealt{Dwarkadas}). For example, \cite{2007ApJ...667..226D} suggests that stellar winds of massive progenitors produce low-denisty cavities in the CSM where SNRs are called to evolve. The front shells of these cavities can be massive enough that they can interact with the shock front of the SNR through reflection shocks. This can cause the bypass, e.g., of the Sedov-Taylor phase and, hence, confine the SNR evolution. Furthermore, the ejecta of (young) SNRs can significantly interact with the (disturbed) CSM and modify it. 

Therefore, as also theoretically expected (e.g. equations \ref{eq:3} and \ref{eq:4}), SNR evolution depends not only on the progenitor, but also the special characteristics of the environment they expand into. This causes older objects in Fig. \ref{fig:cioffi_curves} to be scattered over a large range of values for both shock velocity and electron density; i.e., the interplay with the CSM and ISM in older objects dilutes the trend suggested by younger objects. The highest velocity objects at ages $>10^{4}$\,yr, in particular, could potentially be objects associated with low-density cavities in the CSM, due to their higher velocities compared to the bulk of objects with similar ages. A larger population of younger objects would help to elucidate this picture by providing information on the evolution of SNRs still expanding within the wind cavity or soon after reaching the cavity front. 

This framework is not accounted for by the model of \cite{Cioffi} which assumes a uniform ambient medium. In fact, the only dependence on the density during the radiative phase is through the shock velocity and time at the transition from the Sedov-Taylor to the radiative phase. This determines the initial velocity and time after which the shock velocity evolves as $v_{rad}\propto t^{-7/10}$ (Fig. \ref{fig:cioffi_curves}). In reality, the evolution of a SNR depends strongly on the path it took to get to its current state; i.e. the SNR has "memory" of the past interactions with the medium it expands into. Consequently, the model of \cite{Cioffi} is only a first-order approximation of a more complex process and can only be evaluated as such.

Therefore, one would expect that an SNR during its evolution would cross lines in Fig. \ref{fig:cioffi_curves}. The temporal trajectory of a single object on this plot would start from the upper left edge following a theoretical line of high ambient density and would steadily progress into lines of lower ambient density, as the shock front surpasses the supernova ejecta and CSM and expands into the ISM. Indeed, the color-coding scheme in Fig. \ref{fig:cioffi_curves} reveals a trend of decreasing density with increasing age, while young SNRs have densities similar to the lines they fall on. Exceptions may refer to objects with measurements that are either not well sampled or they are from regions that are not representative of the entire remnant. The wide range of the ISM densities is reflected in the fact that older objects are spread out over multiple lines of ambient density.

\subsection{Limitations of this analysis and future directions}
This work is based on a meta-analysis of an extensive review of published data. Although it is the first comprehensive study of the physical parameters of Galactic SNRs, it is subject to the limitations of the published data. These limitations include:
\begin{enumerate}
\item  The inherent biases in the sample of Galactic SNRs, and mainly the lack of young SNRs that do not allow us to sample adequately their properties and explore their evolution. This can be remedied to some extent by including in future works of this analysis objects from the Magellanic Clouds or even more distant galaxies.

\item The lack of measurements for several objects. This work focused on optical data which is the majority of consistently measured densities and shock velocities for the same regions. However, many objects lack such measurements (c.f. Figs. \ref{fig:den_per_obj}, \ref{fig:vel_per_obj}). An optical investigation on the entire Galactic SNR population is in progress, aiming to map the properties of each SNR (Leonidaki et al., to be submitted). This campaign covers hitherto unknown areas in order to close the existing observational gap and provide all the necessary, missing information. Furthermore, many of these measurements (especially in the case of shock velocities) are based on different methods which often have systematic offsets. Different methods are used depending on the observability or absence of emission lines. In our analysis we opted to treat these measurements independently. Uniform methods would be ideal for consistency, but are not always feasible because, due to the different physical conditions in the SNRs, the necessary emission lines are not available (at sufficient S/N) for all remnants. While we could expand our database  by considering measurements from other wavebands (e.g. radio, X-ray) these would further increase the biases between different methods since other wavebands trace different phases of the SNR material, so this analysis is deferred for a future extension of this work.

\item The lack of intra-object measurements for the majority of objects limits our analysis of the dispersion of physical conditions within different objects to a few well studied cases. The availability of integral field spectroscopy (e.g. from the SDSS-V) survey will significantly help in this front.
\end{enumerate} 

Nonetheless, despite these limitations, this work provides important insights on the physical conditions in SNRs and their dependence on the SNR age and environment.

%% file: sections/conclusions.tex
\section{Conclusions}
\label{sec:conclusions}
Over 300 SNRs are known in our Galaxy. However, so far there have not been any systematic studies of their population (mostly in the optical band). We present an investigation of the physical properties, namely shock velocity and electron density, of Galactic SNRs based on an extensive literature survey. We explore the correlations between velocity and density of the SNRs as a function of their age, providing for the first time a picture of the overall trends of the properties of the SNR population within our Galaxy. More specifically, in this work:
\begin{enumerate}
  \item We thoroughly examined all available publications of the entire SNR sample focusing on measurements of their shock or expansion velocity, density and temperature, based on a variety of methods and tracers in the optical band. We found data for 64 SNRs with published information. For 34 objects, we also have information on multiple regions within the SNR, but we only present results for 9 of those with a statistically sufficient number of regions sampled, providing a picture of the variation of the physical parameters within an object by calculating their standard deviations.

  \item We developed our own scheme for handling and combining various data types, including value ranges, approximate values and upper and lower limits, among regular values with errors, by employing a Monte Carlo method where we drew values from appropriate probability distributions for each one of the measurements. Overall, this probabilistic methodology allowed us to effectively deal with various types of special data, taking into account their inherent uncertainties and constraints. The approach presented here provides a robust framework for handling and analyzing all available data from complex datasets, enabling more complete and representative results.

  \item We find that the density of the SNR population follows a log-normal distribution with a mean $\log(n_e)=2.29$ (cm$^{-3}$) and a standard deviation of $0.49$. The shock velocity of the population follows a skew log-normal distribution located at $\log(\nu_{sh.})=1.59$ (km s$^{-1}$), with a skewness of $6.24$ and a scale of $0.84$. 

  \item Analysis of the intra-object properties for the objects with adequate sample of regions shows that younger SNRs exhibit larger dispersion in electron density and shock velocity while older SNRs exhibit smaller scatter. This is consistent with a picture where young SNRs expand in a clumpy medium while older objects expand in more uniform media.
  
  \item We also explore the relation between shock velocity and density or age. We see that there is a weak positive correlation between shock velocity and density which is driven by the highest velocity SNRs expanding in the densest environments. These also tend to be the youngest ones. However, there is a clear anticorrelation between velocity and age which is mainly driven by the youngest objects, although there is a trend with significant scatter for the older objects ($>3\times10^{3}$\,yrs). This suggests that shock velocities of very young SNRs seem to be driven by age. This means that their kinetic energy, when they are young, is not significantly affected by their ambient medium, even in dense environments. For older SNRs shock velocities are more affected by their surrounding density, resulting in increased scatter. Comparison with evolutionary models shows remarkable agreement, even when considering basic models that do not include all the complexity in the conditions of the CSM and the ISM. The clumpiness of the CSM and/or the initial expansion of the SNR in a stellar-wind blown cavity explains the observed scatter in the shock-velocity - age correlation.

  \item Our analysis suggests that while in the younger SNRs there is stronger intra-object dispersion of the shock velocity (reflecting e.g. the presence of clumps, cavities etc.) there is weak dependence of their average shock velocity on their average density, and stronger dependence of their average shock velocity on their age. Age dependence is also evident in the shock velocity of the older SNRs, but with larger scatter than in the younger ones. Therefore, the overall distribution of the shock velocity is driven by the SNR age, while density plays a secondary role, mostly in increasing the scatter.
\end{enumerate}

The work presented in this study is of great value as it offers a first glimpse into the properties of Galactic SNRs as a population, and a systematic comparison with theoretical models. For instance, the statistical distributions of parameters such as density and shock velocity derived here can be used as inputs in theoretical models that can be constructed to predict other parameters (e.g. \citealt{Kopsacheili2022}), providing a roadmap for understanding the behavior of individual remnants and their collective impact within our Galaxy. 

While our study effectively provides a first picture of the overall trends of the properties of the optical SNR population within our Galaxy and supports our current understanding of shock formation and propagation, it should be noted that this depiction is somewhat rudimentary. The primary challenge we encountered was the scarcity and limited robustness of available data. Notably, temperature data largely relied on theoretical shock models rather than direct measurements, limiting our ability to explore its correlation to other parameters. As a result, our study refrained from investigating potential relationships between temperature and other SNR characteristics. Additionally, age estimates were often highly uncertain. 

Future research endeavors should prioritize the construction of a more extensive and precise dataset, ideally complemented by firsthand spectroscopic observations across multiple regions of SNRs. While this approach may demand additional time and resources for deducing physical properties directly from observations, it opens the door to comprehensive cross-correlations involving shock velocity, electron density, temperature, excitation parameters (e.g. emission line ratios), and supernova progenitors. Such investigations (e.g. based on the SDSS-V Milky-Way Mapper data) could yield valuable insights into the nature of these phenomena, including variations in density and shock velocity associated with different supernova types.

%% file: tables/table.tex
\begin{xltabular}{1.34\textwidth}{p{0.03\textwidth}Xp{0.04\textwidth}p{0.04\textwidth}XccXX}
\caption{Basic properties of the SNRs used in our study. \emph{l} and \emph{b} are the Galactic latitude and longitude, respectively. Multiple age and distance estimates are given from different studies and measurement methods used. A lot of these estimates are reported in the High-Energy Catalogue of Supernova Remnants of \protect\cite{Safi-Harb}. Summary statistics ($\upsilon_{sh.}$, $n_e$) from our analysis, including measurements from different publications and/or regions within a remnant, are provided for the physical parameters. Cases where the standard deviation is zero are associated with objects for which there are only a few and certain measurements. The number of publications based on optical observations per object used in this work are also shown.}
\label{tab:table} \\

\hline\hline 

\multicolumn{1}{l}{\large No. \Large \phantom{N}} & 
\multicolumn{1}{l}{\large SNR Name(s)} &
\multicolumn{1}{l}{\large l} &
\multicolumn{1}{l}{\large b} &
\multicolumn{1}{l}{\large Age} &
\multicolumn{1}{l}{\large $\langle\upsilon_{sh.}\rangle\pm\sigma$} &
\multicolumn{1}{l}{\large $\langle n_{e}\rangle\pm\sigma$} &
\multicolumn{1}{l}{\large Distance} &
\multicolumn{1}{l}{\large No. of Pubs.} \\

\multicolumn{1}{l}{} &
\multicolumn{1}{l}{} &
\multicolumn{1}{l}{(deg.)} &
\multicolumn{1}{l}{(deg.)} &
\multicolumn{1}{l}{(yrs)} &
\multicolumn{1}{c}{(km s$^{-1}$)} &
\multicolumn{1}{c}{(cm$^{-3}$)} &
\multicolumn{1}{l}{(kpc)} &
\multicolumn{1}{l}{}

\endfirsthead

\hline
1 & Kepler & 4.5 & +6.8 & 416 & 1679±413 & 18129±16294 & 2.9±0.4 & 2$^{(a),\,(b)}$ \\
2 & W28 & 6.4 & −0.1 & 33000-36000 & 48±25 & 91±35 & 2, 1.8-3.55 & 2$^{(c),\,(d)}$ \\
3 & 1814-24 & 7.7 & -3.7 & 1200±600 & - & 250±153 & 4.5±1.5 & 1$^{(db)}$ \\
4 & G13.3−1.3 & 13.3 & −1.3 & - & 60±0 & - & 3.3±1.3 & 1$^{(e)}$ \\
5 & G15.1−1.6 & 15.1 & −1.6 & - & 100±0 & 125±41 & 2.1-2.2 & 1$^{(f)}$ \\
6 & G17.4−2.3 & 17.4 & −2.3 & - & 220±153 & 240±51 & - & 1$^{(g)}$ \\
7 & Kes 78 & 32.8 & −0.1 & 5700-22000 & 49±29 & 100±58 & 6-8.5 & 1$^{(h)}$ \\
8 & W44 & 34.7 & −0.4 & 6400-7500 & 130±19 & 497±788 & 2.1-3.3, 2.5 & 2$^{(i),\,(j)}$ \\
9 & G38.7 -1.3 & 38.7 & −1.3 & 3800-14700 & 237±93 & - & - & 1$^{(k)}$ \\
10 & W50 & 39.7 & −2.0 & 30000-100000, 18000-210000 & 91±33 & 1238±1096 & 5, 2-6 & 2$^{(l),\,(m)}$ \\
11 & W51C & 49.2 & −0.7 & $\sim$30000 & 500±107 & 112±78 & $\sim$6 & 1$^{(n),\,(da)}$ \\
12 & 3C 400.2 & 53.6 & −2.2 & 15000-50700, 110000 & 144±128 & 163±207 & 6.7±0.6, 6.7-7.8 & 5$^{(o),\,(p),\,(q),\,(r),\,(s)}$ \\
13 & HC40 & 54.4 & −0.3 & 61000 & 45±26 & 261±247 & ≲3 & 1$^{(t)}$ \\
14 & G59.5+0.1 & 59.5 & +0.1 & - & 90±9 & 449±136 & 11 & 1$^{(u)}$ \\
15 & G59.8+1.2 & 59.8 & +1.2 & - & - & 99±22 & - & 1$^{(t)}$ \\
16 & G64.5+0.9 & 64.5 & +0.9 & - & 143±102 & 272±814 & $\sim$11 & 1$^{(v)}$ \\
17 & G65.3+5.7 & 65.3 & +5.7 & 20000-30000 & 614±338 & - & ≈1.2, 0.6-1.5, 1 & 5$^{(w),\,(x),\,(y),\,(z),\,(aa)}$ \\
18 & G66.0−0.0 & 66.0 & −0.0 & - & 240±93 & - & 2.3-3.96 & 1$^{(k)}$ \\
19 & G67.6+0.9 & 67.6 & +0.9 & - & 80±13 & 147±58 & - & 1$^{(k)}$ \\
20 & G67.7+1.8 & 67.7 & +1.8 & 1500-13000 & - & 621±63 & 7-17, 16.7 & 2$^{(u),\,(n)}$ \\
21 & G67.8+0.5 & 67.8 & +0.5 & - & 94±38 & 102±71 & - & 1$^{(k)}$ \\
22 & CTB 80 & 69.0 & +2.7 & $\sim$10000, 60000, 30000 & 51±29 & 198±162 & 2.5, 1.5-4.6 & 4$^{(ab),\,(ac),\,(ad),\,(ae)}$ \\
23 & G70.0−21.5 & 70.0 & −21.5 & - & 45±26 & 26±15 & 1-2 & 2$^{(af),\,(ag)}$ \\
24 & G73.9+0.9 & 73.9 & +0.9 & 11000-12000 & 222±96 & 558±766 & - & 1$^{(ah)}$ \\
25 & Cygnus Loop & 74.0 & −8.5 & 18000, $\sim$10000 & 30±30 & 689±151 & $\sim$0.89 & 7$^{(ai),\,(aj),\,(ak),\,(al),\,(am),\,(an),\,(ao),\,(cy),\,(cz)}$ \\
26 & DR4, $\gamma$ Cygni SNR & 78.2 & +2.1 & 8000-16000, $\sim$7000 & 174±150 & 57±67 & - & 3$^{(d),\,(ap),\,(aq)}$ \\
27 & W63 & 82.2 & +5.3 & 13500-26700 & 80±0 & 397±85 & 1.6-3.3 & 2$^{(ar),\,(as)}$ \\
28 & G85.9−0.6 & 85.9 & −0.6 & 6400-49000 & 50±29 & 141±102 & 5 & 1$^{(at)}$ \\
29 & HB21 & 89.0 & +4.7 & 4800-18000 & 532±61 & 611±179 & 0.8-1.7 & 1$^{(au)}$ \\
30 & CTB 109 & 109.1 & −1.0 & 8800-14000, 9000-9200 & 3029±2610 & 3548±3817 & 3.6-5.2, 3.1±0.2 & 2$^{(av),\,(aw)}$                            \\
31 & Cas A & 111.7 & −2.1 & 340 & 237±148 & 209±138 & 3.3±0.1, 3.4 & 3$^{(ax),\,(ay),\,(az)}$ \\
32 & G114.3+0.3 & 114.3 & +0.3 & $\sim$41000 & 95±21 & 89±52 & 2-3 & 1$^{(ba)}$ \\
33 & G116.5+1.1 & 116.5 & +1.1 & 15000-50000 & 86±12 & 50±29 & $\sim$3 & 1$^{(bb)}$ \\
34 & CTB 1 & 116.9 & +0.2 & 7500-18100, 7500-11000, 16000 & 178±148 & 78±48 & 1-4.7, 0.9-4.7, 2-3.5 & 3$^{(y),\,(bc),\,(bd)}$                            \\
35 & CTA 1 & 119.5 & +10.2 & 13000 & 146±61 & 477±277 & 1.4, 1.1-1.7 & 2$^{(be),\,(bf)}$ \\
36 & Tycho & 120.1 & +1.4 & 451 & 2428±662 & - & - & 2$^{(bg),\,(bh)}$ \\
37 & G126.2+1.6 & 126.2 & +1.6 & 270000 & 110±23 & 312±236 & 4.5, 2-5 & 2$^{(t),\,(bi)}$ \\
38 & 3C58 & 130.7 & +3.1 & 839 & 664±242 & 2621±3289 & 2.6±0.2 & 2$^{(bj),\,(bk)}$ \\
39 & HB3 & 132.7 & +1.3 & 25000-72000 & 52±29 & 75±43 & 2-2.2 & 2$^{(ap),\,(bl)}$ \\
40 & G156.2+5.7 & 156.2 & +5.7 & 7000-36600, ≳20000 & 100±25 & 268±57 & 0.68-3, ≳1.7 & 2$^{(bm),\,(bn)}$ \\
41 & G159.6+7.3 & 159.6 & +7.3 & - & 605±232 & - & \textless{}2.5 & 1$^{(bo)}$ \\
42 & VRO 42.05.01 & 166.0 & +4.3 & 9000-20100, 60000 & 82±24 & 149±86 & 2-3.6 & 2$^{(y),\,(bp)}$ \\
43 & G179.0+2.6 & 179.0 & +2.6 & \textgreater{}10000 & 297±115 & 50±29 & $\sim$3.5 & 1$^{(bq)}$ \\
44 & S147 & 180.0 & −1.7 & 26000-34000 & 101±16 & 119±70 & 0.8-0.9, 0.6-1.9 & 3$^{(y),\,(ap),\,(br)}$ \\
45 & Crab Nebula & 184.6 & −5.8 & 966 & 875±640 & 1131±773 & - & 2$^{(bs),\,(bt)}$ \\
46 & IC443, 3C157 & 189.1 & +3.0 & 9000, $\sim$10000 & 217±167 & 433±274 & 0.5-2.5 & 2$^{(bp),\,(bu)}$ \\
47 & Monoceros Nebula & 205.5 & +0.5 & 30000-150000 & 44±26 & 150±86 & - & 2$^{(y),\,(ap)}$ \\
48 & PKS 0646+06 & 206.9 & +2.3 & 64000 & 178±69 & 800±0 & 3-6.5, 1-2.3, $\sim$2.2 & 3$^{(y),\,(bv),\,(bw)}$ \\
49 & G213.0−0.6 & 213.0 & −0.6 & - & - & 100±21 & $\sim$2.4 & 1$^{(bx)}$ \\
50 & Puppis A, MSH 08−44 & 260.4 & −3.4 & 2200-5400 & 2236±639 & - & 2-4 & 1$^{(by)}$ \\
51 & Vela & 263.9 & −3.3 & 9000-27000 & 106±56 & 6±2 & 0.25 & 2$^{(bz),\,(ca)}$ \\
52 & MSH 10−53 & 284.3 & −1.8 & $\sim$10000, 2930-3050 & 90±0 & - & 1-2.9, 3.7-5.4, 6-6.2, 6.2±0.9 & 1$^{(cb)}$                            \\
53 & MSH 11−54 & 292.0 & +1.8 & 2930-3050, 2700-3700 & - & 250±249 & 3.7-5.4, 6-6.2, 6.2±0.9 & 3$^{(cc),\,(cd),\,(ce)}$                            \\
54 & G296.1−0.5 & 296.1 & −0.5 & 2800-28000 & - & 252±150 & 3-5 & 1$^{(cf)}$ \\
55 & Milne 23, PKS 1209−51/52 & 296.5 & +10.0 & 7000-10000 & - & 5±0 & 1.3-3.9 & 1$^{(cg)}$ \\
56 & G299.2−2.9 & 299.2 & −2.9 & ≈8700 & 498±108 & - & ≈5 & 1$^{(ch)}$ \\
57 & G315.1+2.7 & 315.1 & +2.7 & - & 99±21 & 33±20 & 1.7 & 1$^{(ci)}$ \\
58 & RCW 86, MSH 14−63 & 315.4 & −2.3 & 2000-12400 & 491±331 & 1012±262 & 2.5±0.5, 2.3±0.2, 3.2 & 5$^{(b),\,(bg),\,(cj),\,(ck),\,(cl)}$                            \\
59 & MSH 15−52, RCW 89 & 320.4 & −1.2 & 1700-1900 & - & 15±9 & 4 & 1$^{(cm)}$ \\
60 & MSH 15−56 & 326.3 & −1.8 & 9800-16500 & - & 360±75 & 3.2 & 1$^{(cn)}$ \\
61 & SN1006, PKS 1459−41 & 327.6 & +14.6 & 1017 & 2941±1052 & - & 2.1 & 5$^{(co),\,(cp),\,(cq),\,(cr),\,(cs)}$ \\
62 & RCW 103 & 332.4 & −0.4 & 2000, 2000-4400, 1200-3200 & 655±1144 & 1352±1406 & 3.3, 3.2, 2.7-3.3 & 4$^{(b),\,(cg),\,(ct),\,(cu)}$                            \\
63 & G332.5−5.6 & 332.5 & −5.6 & 7000-12100 & - & 297±248 & 2.2-3.8 & 1$^{(cv)}$ \\
64 & RCW 114 & 343.0 & −6.0 & $\sim$20000 & 30±4 & 77±67 & 0.2-1.5 & 1$^{(cw)}$ \\ \hline
\caption*{\justifying The references to the publications we have retrieved our data from can be found \protect\hyperlink{references}{here}.}
\end{xltabular}

%% file: appendices/appendix_b.tex
\normalsize
\justifying
The table below provides a comprehensive overview of the properties of our sample of Galactic SNRs. It lists all the SNRs in our sample along with specific regions within them, for which we have shock velocity, electron density, and temperature measurements. These data are crucial in elucidating the diverse physical characteristics and dynamical processes within these objects that we have explored in this work, aiding in a deeper understanding of their complex structures and interactions with the interstellar medium.

%% file: tables/dataset.tex
\begin{xltabular}{1.34\textwidth}{XXXcccccc}
\caption{SNR properties}
\label{tab:dataset} \\
\hline\hline
Object / Name(s) \phantom{\Large I} & Region & \multicolumn{2}{c}{Shock Velocity} & \multicolumn{2}{c}{Electron Density} & Temperature & References \\ 
 & & \multicolumn{2}{c}{(km s$^{-1}$)} & \multicolumn{2}{c}{(cm$^{-3}$)} & (K) & \\ \hline
 & & Measurement & Method & Post-shock & Pre-shock & & \\ \hline
G4.5+6.8 / Kepler & Knot D3 & 1550-2000 & Line Broadening & 2100±100 & $\sim$100 & 22300±5100 & a, b \\
 & Knot D9 & 1550-2000 & Line Broadening & 4900 & $\sim$100 & 41000±24000 & \\
 & Knot D18 & 1550-2000 & Line Broadening & 5000±500 & $\sim$100 & 28200±1500 & \\
 & Knot D34, 35 & 1550-2000 & Line Broadening & 5400 & $\sim$100 & - & \\
 & Knot D38, 40 & 1550-2000 & Line Broadening & 4000 & $\sim$100 & - & \\
 & Knot D63, 64 & 1550-2000 & Line Broadening & $\ge$10000 & $\sim$100 & - & \\
 & Knot D41-45 & 1550-2000 & Line Broadening & <100 & $\sim$100 & 17000 & \\
 & Knot D55 & 1550-2000 & Line Broadening & $\ge$10000 & $\sim$100 & 17200 & \\
 & Knot D56 & 1550-2000 & Line Broadening & $\ge$10000 & $\sim$100 & - & \\
 & Knot D61 & 1550-2000 & Line Broadening & $\ge$10000 & $\sim$100 & - & \\
 & SW of D9 & 1550-2000 & Line Broadening & 5400 & $\sim$100 & - & \\
 & Knot D9, 10 & 1550-2000 & Line Broadening & 8000 & $\sim$100 & - & \\
 & Knot D25 & 1550-2000 & Line Broadening & 5800 & $\sim$100 & 21600 & \\
 & Knot D27 & 1550-2000 & Line Broadening & 7000 & $\sim$100 & 43000 & \\
 & [S II] Regions & $\ge$100 & Line Ratios & 5000 & >1000 & 10000-15000 & \\ \hline
G6.4−0.1 / W28 & I (NE) & ≲70 & Line Ratios & 70±50 & - & - & c, d \\
 & II (S) & ≲70 & Line Ratios & 110±50 & - & - & \\
 & III (NW) & ≲70 & Line Ratios & 100±20 & - & - & \\
 & IV (CW) & 40-50 & Line Ratios & 100±30 & - & - & \\
 & IVa (CW) & ≲70 & Line Ratios & 90±20 & - & - & \\
 & V (CE) & ≲70 & Line Ratios & 100±20 & - & - & \\
 & Va (CE) & ≲70 & Line Ratios & 60±50/40 & - & - & \\
 & R1-1 & 60-90 & Line Ratios & 110 & - & $\sim$10000\ensuremath{^a} & \\
 & R1-2 & 85 & Line Ratios & 275 & - & $\sim$10000\ensuremath{^a} & \\
 & R1-3 & 60-90 & Line Ratios & 100 & - & $\sim$10000\ensuremath{^a} & \\
 & R2 & 60-90 & Line Ratios & 300 & - & $\sim$10000\ensuremath{^a} & \\ \hline
G7.7−3.7 / 1814-24 & Filament B Pos. 1 & - & - & <400 & - & 10000\ensuremath{^a} & db \\
 & Filament B Pos. 2 & - & - & <600 & - & 10000\ensuremath{^a} & \\
G13.3−1.3 & Filaments (S, NE) & 60 & Line Ratios & - & - & >50000 & e \\
G15.1−1.6 & North I & 100 & Line Ratios & 170 & - & $\sim$10000\ensuremath{^a} & f \\
 & North II & 100 & Line Ratios & 134 & - & $\sim$10000\ensuremath{^a} & \\
 & SE & 100 & Line Ratios & 70 & - & $\sim$10000\ensuremath{^a} & \\ \hline
G17.4−2.3 & Optical filaments & >100 & Line Ratios & $\sim$240 & - & $\sim$10000\ensuremath{^a} & g \\
 & E filament, Position Ia & >100 & Line Ratios & 290 & - & $\sim$10000\ensuremath{^a} & \\
 & E filament, Position Ib & ≲100 & Line Ratios & 50 & - & $\sim$10000\ensuremath{^a} & \\ \hline
G32.8−0.1 / Kes 78 & South I & $\le$100 & Line Ratios & $\le$200 & - & 10000\ensuremath{^a} & h \\
 & South II & $\le$100 & Line Ratios & $\le$200 & - & 10000\ensuremath{^a} & \\
 & East I & $\le$100 & Line Ratios & $\le$200 & - & 10000\ensuremath{^a} & \\
 & East II & $\le$100 & Line Ratios & $\le$200 & - & 10000\ensuremath{^a} & \\
 & Optical filaments (S, E) & $\le$100 & Line Ratios & $\le$200 & - & 10000\ensuremath{^a} & \\ \hline
G34.7−0.4 / W44 & Entire SNR (Average) & 630 & Other & 900-2600 & 0.09-0.26 & 2700000-8600000 & i, j \\
 & Area I & 110-150 & Line Ratios & <220 & - & 2.7 & \\
 & Area II & 110-150 & Line Ratios & <140 & - & 6.2 & \\
 & Area III & 110-150 & Line Ratios & <200 & - & - & \\ \hline
G38.7−1.3 & Slit position 1 & >80 & Line Ratios & - & - & - & k \\
G39.7−2.0 / W50 & Filaments (E) & $\simeq$100 & Line Ratios & 50 & - & - & l, m \\
 & Filaments (W) & $\simeq$120 & Line Ratios & 700 & - & - & \\
 & West I & 40-106 & Line Ratios & 3000 & $\sim$300 & - & \\
 & West II & 40-106 & Line Ratios & 1200 & $\sim$120 & - & \\ \hline
G49.2−0.7 / W51C & Shock front & $\sim$500 & Line Ratios & <240 & - & $\sim$3000000 & n, da \\
& Cloudlets & 75 & Line Ratios & 1-200 & - & 10000\ensuremath{^a} & \\
G53.6−2.2 / 3C 400.2 & Entire SNR (Average) & >100 & Line Ratios & 900 & - & - & o, p, q, r, s \\
 & Optical filaments & $\simeq$60 & Doppler & 96-796 & 4 & - & \\
 & Slit position 1 & - & - & 30-75 & - & 10000\ensuremath{^a} & \\
 & Slit position 2 & - & - & 140-300 & - & 10000\ensuremath{^a} & \\
 & Optical shell & - & - & ≲70 & - & 10000\ensuremath{^a} & \\ \hline
G54.4−0.3 / HC40 & Entire SNR (Average) & $\le$90 & Line Ratios & <50 & - & - & t \\
 & Mean of Area Ia-c \& Area II & - & - & ≈500 & - & - & \\ \hline
G59.5+0.1 & Area 1 & 80 & Line Ratios & 313 & 11 & $\sim$10000\ensuremath{^a} & u \\
 & Area 2 & 80-100 & Line Ratios & 585 & 16 & $\sim$10000\ensuremath{^a} & \\ \hline
G64.5+0.9 & Optical filaments (N, W) & - & - & $\sim$100 & - & - & v \\
G65.3+5.7 & O[III] filaments & $\ge$50 & Line Ratios & - & ≈0.5 & - & w, x, y, z, aa \\
 & $\phi$ H$\alpha$ filaments & 100±30 & Doppler & $\gtrapprox$1000 & >2.5 & - & \\
 & Bright [OIII] filament (Position 2) & - & - & <300 & - & 38000±7000 & \\
 & Areas of shock heated gas & 90-140 & Line Ratios & ≲200 & - & - & \\
 & Area 1a & 120 & Line Ratios & <150 & - & - & \\
 & Area 1b & 120 & Line Ratios & <140 & - & - & \\
 & Area 1c & 120 & Line Ratios & <140 & - & - & \\
 & Area 2 & 90-140 & Line Ratios & <130 & - & - & \\
 & Area 3a & 90-140 & Line Ratios & <45 & - & - & \\
 & Area 3b & 90-140 & Line Ratios & <70 & - & - & \\
 & Area 4 & 90-140 & Line Ratios & <35 & - & - & \\
 & Area 5 & 90 & Line Ratios & <30 & - & - & \\
 & Area 6 & 90-140 & Line Ratios & <30 & - & - & \\
 & Area 7 & 90-140 & Line Ratios & <170 & - & - & \\
 & Recombination zone  & 90-140 & Other & 30-170 & - & $\sim$10000\ensuremath{^a} & \\
 & Entire SNR (Average) & 400±200\ensuremath{^b} & Other & - & - & - & \\ \hline
G66.0−0.0 & Optical filaments & 200-1000 & Line Ratios & - & - & - & k \\
G67.6+0.9 & Optical filaments & >80 & Line Ratios & - & - & - & k \\
G67.7+1.8 & Area 1 & 80 & Line Ratios & 358 & 12 & $\sim$10000\ensuremath{^a} & u,n \\
 & Area 2 & 80-100 & Line Ratios & 480 & 13 & $\sim$10000\ensuremath{^a} & \\
 & Entire SNR & 70 & Line Ratios & $\sim$142 & - & $\sim$10000\ensuremath{^a} & \\
 & Optical filament (SW-NE) & 60-80 & Line Ratios & 60-240 & - & - & \\ \hline
G67.8+0.5 & Entire SNR (Average) & - & - & 620±74 & - & - & k \\
G69.0+2.7 / CTB 80 & SW & $\le$90 & Line Ratios & 200 & - & $\simeq$10000\ensuremath{^a} & ab, ac, ad, ae \\
 & Cross-section 126-132 of slit Pos. 2 & $\simeq$120 & Line Ratios & 166±47 & - & $\simeq$10000\ensuremath{^a} & \\
 & Rest of slit Pos. 2 & $\simeq$120 & Line Ratios & $\le$80 & - & $\simeq$10000\ensuremath{^a} & \\
 & Radiative shocks in core & 120-140 & Line Ratios & - & 50-100 & - & \\
 & Filaments (Central) & 80-100 & Line Ratios & - & 50-100 & - & \\
 & Position A1 & - & - & 141 & 50-100 & 9700 & \\
 & Position A2 & - & - & 225 & 50-100 & 14900 & \\
 & Position A3 & - & - & 141 & 50-100 & 12100 & \\
 & Position B1 & - & - & - & 50-100 & - & \\
 & Position B2 & - & - & 141 & 50-100 & 11400 & \\
 & Position B3 & - & - & 141 & 50-100 & 13200 & \\
 & Position B4 & - & - & 195 & 50-100 & 11600 & \\
 & Position C1 & - & - & 225 & 50-100 & 14100 & \\
 & Position C2 & - & - & 250 & 50-100 & 13000 & \\
 & Position D1 & - & - & 109 & 50-100 & - & \\
 & Position D2 & - & - & - & 50-100 & - & \\
 & SW Area  & 85-120 & Line Ratios & - & 2.6-5.2 & - & \\
 & East Area  & <100 & Line Ratios & - & <5 & - & \\ \hline
G70.0−21.5 & Optical filaments & $\le$100 & Line Ratios & <600 & - & - & af, ag \\
 & Filament 2 & <100 & Line Ratios & $\le$200 & - & 10000\ensuremath{^a} & \\ \hline
G73.9+0.9 & Recombination zone  & <90 & Line Ratios & <50 & - & - & ah \\
G74.0−8.5 / Cygnus Loop & NW Limb & 175-185 & Line Ratios & - & 5-12 & 500000 & ai, aj, ak, al, am, an, ao, cy, cz \\
 & NW Filament 2 (SE End) & $\sim$180 & Line Ratios & - & - & - & \\
 & NW Filament 2 (NW End) & $\sim$140 & Line Ratios & - & - & - & \\
 & Optical filaments NE0, NE1, E0, E1 and SW0 & - & - & $\le$50 & - & - & \\
 & Face-on H$\alpha$ emission near center  & 350-400 & Other & - & - & - & \\
 & Bright, radiative filaments & $\sim$100 & Other & - & - & - & \\
 & Faint, partially radiative/nonradiative filaments & 150-200 & Other & - & - & - & \\
 & P7 filament & 140-400 & Line Ratios & 200-2000 & - & 1000-10000 & \\
 & Entire SNR (Average) & $\sim$360 & Other & - & - & - & \\
 & Balmer-dominated shock in NW limb & $\sim$200 & Other & - & - & - & \\
 & S10 fiber in optical filament  & 254 & Line Broadening & - & - & 1600000 & \\
 & S89 ON fiber in optical filament  & 294 & Line Broadening & - & - & 2900000 & \\
 & S89 OFF fiber in optical filament  & 294 & Line Broadening & - & - & 2100000 & \\
 & S7 fiber in optical filament  & 240 & Line Broadening & - & - & 1800000 & \\
 & S6 fiber in optical filament  & 333 & Line Broadening & - & - & 2100000 & \\
 & S5 fiber in optical filament  & 278 & Line Broadening & - & - & 1900000 & \\
 & S4 fiber in optical filament  & 225 & Line Broadening & - & - & 2000000 & \\
 & Region 1  & 190-290 & Other & - & 0.95-2.2 & - & \\ 
 & Region 2  & 155 & Other & - & 3.4 & - & \\
 & Eastern filaments  & 150-290 & Other & - & $\sim$5 & $\sim$15000 & \\ \hline
G78.2+2.1 / DR4, $\gamma$ Cygni SNR & Entire SNR (Average) & <15\ensuremath{^b} & Doppler & 65 & - & 10000\ensuremath{^a} & d, ap, aq \\
 & Region 1 & - & - & 125 & - & - & \\
 & Region 2 & - & - & 75 & - & - & \\
 & Region 3 & - & - & 10 & - & - & \\
 & Optical filament & <100 & Line Ratios & $\sim$700 & $\sim$20 & - & \\ \hline
G82.2+5.3 / W63 & Cygnus X & - & - & 100-220 & - & 10000\ensuremath{^a} & ar, as \\
 & Area I (West) & <100 & Line Ratios & <80 & - & - & \\
 & Area II (East) & >100 & Line Ratios & <30 & - & - & \\
 & Area III (South) & - & - & <30 & - & - & \\ \hline
G85.9−0.6 & Average physical parameters of shell & 80 & Line Ratios & 395 & 14 & $\sim$10000\ensuremath{^a} & at \\
 & Area 1 & - & - & 470±120 & 16±3 & $\sim$10000\ensuremath{^a} & \\
 & Area 2 & - & - & 380±40 & 13±1 & $\sim$10000\ensuremath{^a} & \\
 & Area 3 & - & - & 336±26 & 12±1 & $\sim$10000\ensuremath{^a} & \\ \hline
G89.0+4.7 / HB21 & Area I & <100 & Line Ratios & <380 & $\sim$2.5 & - & au \\
 & Area II & <100 & Line Ratios & <120 & $\sim$2.5 & - & \\
 & Area III & <100 & Line Ratios & <130 & $\sim$2.5 & - & \\
 & Area IV & <100 & Line Ratios & <320 & $\sim$2.5 & - & \\
 & Area V & <100 & Line Ratios & <330 & $\sim$2.5 & - & \\
 & Area VI & <100 & Line Ratios & <290 & $\sim$2.5 & - & \\
 & Area VII & <100 & Line Ratios & <410 & $\sim$2.5 & - & \\ \hline
G109.1−1.0 / CTB 109 & Optical filaments & - & - & ≈700 & - & - & av, aw \\
 & Average physical parameters of optical filaments & 460-603 & Doppler & 580±185 & 1.1±0.3 & 5000-10000\ensuremath{^a} & \\
 & NE filaments & 460-603 & Doppler & 501±120 & 1.1±0.3 & 5000\ensuremath{^a} & \\
 & SE filaments & 460-603 & Doppler & 658±250 & 1.1±0.3 & 10000\ensuremath{^a} & \\ \hline
G111.7−2.1 / Cas A & Entire SNR (Average) & 2500-7000\ensuremath{^b} & Doppler & 100-10000 & - & 10000\ensuremath{^a} & ax, ay, az \\
 & Knots & 2500-7000\ensuremath{^b} & Doppler & 1000-10000 & - & 10000\ensuremath{^a} & \\
 & Forward shock & ≈5000 & Doppler & 1000-10000 & - & 10000\ensuremath{^a} & \\
 & Optical knots/clumps & 100-400 & Other & 100-1000 & - & ≈30000 & \\
 & Post-shock gas & 100-500 & Other & ≈1000 & - & 100000-10000000 & \\ \hline
G114.3+0.3 & Slit position I & ≳100 & Line Ratios & ≲270 & - & - & ba \\
 & Slit position II & ≳100 & Line Ratios & ≲600 & - & - & \\
 & Slit position IIIa & ≳100 & Line Ratios & ≲400 & - & - & \\
 & Slit position IIIb & ≲100 & Line Ratios & ≲400 & - & - & \\ \hline
G116.5+1.1 & Slit position VI & 70-120 & Line Ratios & <180 & ≲5 & - & bb \\
G116.6-26.1 & Filament in region UF & 70-100 & Line Ratios & ≲100 & 0.0001 & - & cx \\
G116.9+0.2 / CTB 1 & SW Rim & - & - & 100 & - & - & y, bc, bd \\
 & Western side & $\ge$100 & Line Ratios & <200 & - & - & \\
 & Eastern side & $\sim$80 & Line Ratios & <200 & - & - & \\
 & Main shell & 370 & Line Ratios & $\le$100 & - & 1900000 & \\
 & Western limb optical filaments & ≳100 & Line Ratios & $\le$100 & - & - & \\
 & SE Limb & ≲70 & Line Ratios & $\simeq$90 & - & - & \\ \hline
G119.5+10.2 / CTA 1 & Optical filaments & 100-120 & Line Ratios & <900 & <3 & - & be, bf \\
G120.1+1.4 / Tycho & Knot g & 1200-3000 & Line Broadening & - & 0.3 & - & bg, bh, co \\
 & Entire NE filament & 2500-3000 & Other & - & - & ≈68000 & \\ \hline
G126.2+1.6 & Optical filaments & $\sim$100 & Line Ratios & 125 & ≈3 & 8000-10000\ensuremath{^a} & t, bi \\
 & General postshock physical properties (optical filaments) & 100-120 & Line Ratios & 30-600 & >13.3 & 500000 & \\
 & Area Ia & $\sim$120 & Line Ratios & 30-600 & >13.3 & - & \\
 & Area Ib & $\sim$120 & Line Ratios & 30-600 & >13.3 & - & \\
 & Area II & $\sim$100 & Line Ratios & 30-600 & >13.3 & - & \\ \hline
G130.7+3.1 / 3C58 & Most knots & 770±155\ensuremath{^b} & Line Ratios & 100-500 & 2-10 & - & bj, bk \\
 & Some knots & 770±155\ensuremath{^b} & Line Ratios & 3000-10000 & - & - & \\
 & Central area & ≲900 & Doppler & $\sim$1000 & - & 10000\ensuremath{^a} & \\ \hline
G132.7+1.3 / HB3 & Entire SNR (Average) & - & - & 50 & - & 10000\ensuremath{^a} & ap, bl \\
 & Western limb optical filaments & $\le$100 & Line Ratios & $\le$150 & 0.04-0.06 & - & \\ \hline
G156.2+5.7 & NE1 & - & - & 175 & - & 10000\ensuremath{^a} & bm, bn \\
 & NE2 & - & - & 325 & - & 10000\ensuremath{^a} & \\
 & SW1 & - & - & 270 & - & 10000\ensuremath{^a} & \\
 & SW2 & $\sim$100 & Line Ratios & 300 & 10 & 10000\ensuremath{^a} & \\
 & Optical filaments & 500 & Other & - & - & <3500000\ensuremath{^a} & \\ \hline
G159.6+7.3 & Region 1 of H$\alpha$ shell & ≳200 & Doppler & - & - & - & bo \\
 & Region 2 of H$\alpha$ shell & ≳200 & Doppler & - & - & - & \\ \hline
G166.0+4.3 / VRO 42.05.01 & Northern filament & 60-120 & Line Ratios & <300 & - & - & y, bp \\
 & Entire SNR (Average) & 200 & Doppler & 250 & - & 10000\ensuremath{^a} & \\
 & Filaments \& diffuse regions & 50-100 & Line Broadening & - & - & - & \\ \hline
G179.0+2.6 & Optical filaments (S, NE limbs) & >100 & Line Ratios & <100 & - & - & bq \\
 & SW 2 & >100 & Line Ratios & <100 & - & - & \\
 & NW 1a & >100 & Line Ratios & <100 & - & - & \\
 & Slit position NE 1a & >100 & Line Ratios & $\le$100 & - & - & \\ \hline
G180.0−1.7 / S147 & Slit position 3 & - & - & 100\ensuremath{^a} & - & 52000±10000 & y, ap, br \\
 & Optical filaments & 80-120\ensuremath{^b} & Doppler & 37.5-200 & - & - & \\
 & Entire SNR (Average) & - & - & 250 & - & 10000\ensuremath{^a} & \\ \hline
G184.6−5.8 / Crab Nebula & Entire SNR (Average) & ≈1500\ensuremath{^b} & Other & - & - & - & bs, bt \\
 & Optical filaments (outer regions) & - & - & 40-2700 & - & 11000-18000 & \\
 & SE  & - & - & $\sim$600 & - & 11000-18000 & \\
 & NW region of PWN & $\gg$180 & Doppler & $\sim$1400 & - & 11000-18000 & \\
 & Near the equatorial region of PWN & ≈150 & Doppler & - & - & - & \\ \hline
G189.1+3.0 / IC443, 3C157 & NE sector of optical shell & 200-370 & Doppler & 400-800 & 100-200 & 500000-1500000 & bp, bu \\
 & SW sector of optical shell & 370-530 & Doppler & 400-800 & 100-200 & 1500000-3000000 & \\
 & NE filament (slit position A) & $\sim$70 & Line Ratios & ≈100 & - & ≈9000 & \\
 & Interstellar HI clouds & $\sim$70 & Line Ratios & - & 1-10000 & ≈10000 & \\ \hline
G205.5+0.5 / Monoceros Nebula & Slit position 1 & <90 & Line Ratios & <300 & - & - & y, ap \\
 & Entire SNR (Average) & - & - & 100 & - & 10000\ensuremath{^a} & \\ \hline
G206.9+2.3 / PKS 0646+06 & Optical filaments & >60 & Line Ratios & - & - & - & y, bv, bw \\
 & Slit position 1 & - & - & 800 & - & - & \\ \hline
G213.0−0.6 & Optical filament a) & - & - & $\sim$100 & - & - & bx \\
G260.4−3.4 / Puppis A, MSH 08−44 & Oxygen-rich filaments & 1500-3000\ensuremath{^b} & Other & - & - & 20000-30000\ensuremath{^a} & by \\
G263.9−3.3 / Vela & Optical filaments & 90-110 & Other & 3.2-8.2 & 0.9-2.1 & 30000 - 100000 & bz, ca \\
 & Optical filament in Knot D (Slow shocks) & ≲100 & Line Ratios & - & $\sim$3.5 & 10000\ensuremath{^a} & \\
 & Optical filament in Knot D (Fast shocks) & $\sim$170 & Line Ratios & - & $\simeq$11 & 10000\ensuremath{^a} & \\ \hline
G284.3−1.8 / MSH 10−53 & Optical filament & 90 & Line Ratios & - & 10 & 190000 & cb \\
G292.0+1.8 / MSH 11−54 & Filament & - & - & ≲1000 & - & 36000±1500 & cc, cd, ce \\
 & West & - & - & $\sim$200 & - & - & \\ \hline
G296.1−0.5 & Filament 1 & - & - & 2000 & - & 1000\ensuremath{^a} & cf \\
 & Filament 2 & - & - & <500 & - & 1000\ensuremath{^a} & \\
 & Filament 3 & - & - & <500 & - & 1000\ensuremath{^a} & \\ \hline
G296.5+10.0 / Milne 23, PKS 1209−51/52 & Northern shell & - & - & 5 & - & >250000 & cg \\
G299.2−2.9 & Entire SNR & ≈500\ensuremath{^b} & Line Ratios & - & ≈0.3 & 3500000 & ch \\
G315.1+2.7 & Slit Position A & $\sim$100 & Line Ratios & $\sim$33 & - & 10000\ensuremath{^a} & ci \\
 & Slit Position B & $\sim$100 & Line Ratios & 1-66 & - & 10000\ensuremath{^a} & \\ \hline
G315.4−2.3 / RCW 86, MSH 14−63 & Entire SNR & $\ge$100 & Line Ratios & $\sim$1000 & >100 & 10000\ensuremath{^a} & b, bg, cj, ck, cl \\
 & Position 1 & $\ge$100 & Line Ratios & 1740 & >100 & 10000\ensuremath{^a} & \\
 & Position 2 & $\ge$100 & Line Ratios & 1250 & >100 & 10000\ensuremath{^a} & \\
 & Position 3 & $\ge$100 & Line Ratios & 3420 & >100 & 10000\ensuremath{^a} & \\
 & Position 4 & $\ge$100 & Line Ratios & 5070 & >100 & 10000\ensuremath{^a} & \\
 & Position 5 & $\ge$100 & Line Ratios & 1550 & >100 & 10000\ensuremath{^a} & \\
 & Position 6 & $\ge$100 & Line Ratios & 2380 & >100 & 10000\ensuremath{^a} & \\
 & Position 7 & $\ge$100 & Line Ratios & 1120 & >100 & 10000\ensuremath{^a} & \\
 & Position 8 & $\ge$100 & Line Ratios & 1800 & >100 & 10000\ensuremath{^a} & \\
 & Position 9 & $\ge$100 & Line Ratios & 1820 & >100 & 10000\ensuremath{^a} & \\
 & Position 10 & $\ge$100 & Line Ratios & 1390 & >100 & 10000\ensuremath{^a} & \\
 & Filaments (SW, N) & 500-930 & Line Broadening & - & - & - & \\
 & Filaments (N) & 600±100 & Line Broadening & - & - & - & \\
 & SW & 580-660 & Line Broadening & - & 1 & $\sim$100000\ensuremath{^a} & \\
 & W & 580-660 & Line Broadening & - & - & 20000-60000 & \\
 & NW & 580-660 & Line Broadening & - & - & 20000-60000 & \\
 & NE, SE Regions & 700-2200 & Other & - & - & 26700±3500 & \\ \hline
G320.4−1.2 / MSH 15−52, RCW 89 & Position 1 & - & - & ≲30 & - & - & cm \\
 & Position 2 & - & - & 60 & - & 91000-99000 & \\
 & Position 3 & - & - & 100 & - & 79000-90000 & \\ \hline
G326.3−1.8 / MSH 15−56 & All cases & - & - & $\sim$360 & - & 10000\ensuremath{^a} & cn \\
 & Filament 1 (SE) Pos. 1 & - & - & 340 & - & 10000\ensuremath{^a} & \\
 & Filament 1 (SE) Pos. 2 & - & - & 330 & - & 10000\ensuremath{^a} & \\
 & Filament 2 (NW) & - & - & $\sim$360 & - & 10000\ensuremath{^a} & \\ \hline
G327.6+14.6 / SN1006, PKS 1459−41 & NW Rim & 2800-3870 & Line Broadening & - & - & - & co, cp, cq, cr, cs \\
 & NW Limb & 2890±100 & Line Broadening & - & ≲1 & 5000-10000 & \\
 & Entire NW Rim & 2500-2900 & Line Broadening & - & 0.25-0.4 & - & \\
 & NW Rim Pos 1  & 2500-2900 & Line Broadening & - & >0.35 & - & \\
 & NW Rim Pos 2 & 2500-2900 & Line Broadening & - & >0.4 & - & \\
 & NW Rim Pos 3 & 2500-2900 & Line Broadening & - & 0.3-0.35 & - & \\
 & NW Rim Pos 4 & 2500-2900 & Line Broadening & - & >0.35 & - & \\
 & NW Rim Pos 5 & 2500-2900 & Line Broadening & - & 0.25-0.35 & - & \\
 & NW Rim Pos 6 & 2500-2900 & Line Broadening & - & 0.3-0.4 & - & \\
 & NW Rim Pos 7 & 2500-2900 & Line Broadening & - & 0.25-0.35 & - & \\
 & NW Rim Pos 8 & 2500-2900 & Line Broadening & - & 0.35-0.4 & - & \\
 & Entire SNR & >2000 & Line Broadening & - & - & - & \\
 & NW H$\alpha$ filament & 2500±180 & Line Broadening & - & - & - & \\
 & NW bow shock & 3060±400 & Line Broadening & - & 0.15-0.3 & - & \\ \hline
G332.4−0.4 / RCW 103 & Entire SNR & >100 & Line Ratios & 10-100 & - & $\sim$200000 & b, cg, ct, cu \\
 & I & $\sim$100 & Line Ratios & 1970 & - & $\sim$10000\ensuremath{^a} & \\
 & II & $\sim$100 & Line Ratios & 1750 & - & $\sim$10000\ensuremath{^a} & \\
 & III & $\sim$100 & Line Ratios & 1180 & - & $\sim$10000\ensuremath{^a} & \\
 & IV & $\sim$100 & Line Ratios & 2900 & - & $\sim$10000\ensuremath{^a} & \\
 & V & $\sim$100 & Line Ratios & 2000 & - & $\sim$10000\ensuremath{^a} & \\
 & Optical filaments & 130 & Line Ratios & - & - & - & \\
 & Dense condensation & ≳400 & Other & ≳1000 & - & $\sim$10000\ensuremath{^a} & \\ \hline
G332.5−5.6 & I & - & - & 490 & - & $\sim$10000\ensuremath{^a} & cv \\
 & II & - & - & 740 & - & $\sim$10000\ensuremath{^a} & \\
 & III & - & - & 60 & - & $\sim$10000\ensuremath{^a} & \\
 & IV & - & - & 280 & - & $\sim$10000\ensuremath{^a} & \\
 & V & - & - & 150 & - & $\sim$10000\ensuremath{^a} & \\
 & VI & - & - & 60 & - & $\sim$10000\ensuremath{^a} & \\ \hline
G343.0−6.0 / RCW 114 & I & 25-35\ensuremath{^b} & Line Ratios & 70 & - & $\sim$10000\ensuremath{^a} & cw \\
 & II & 25-35\ensuremath{^b} & Line Ratios & 250 & - & $\sim$10000\ensuremath{^a} & \\
 & III & 25-35\ensuremath{^b} & Line Ratios & 80 & - & $\sim$10000\ensuremath{^a} & \\
 & IV & 25-35\ensuremath{^b} & Line Ratios & 12 & - & $\sim$10000\ensuremath{^a} & \\
 & V & 25-35\ensuremath{^b} & Line Ratios & 4 & - & $\sim$10000\ensuremath{^a} & \\
 & VI & 25-35\ensuremath{^b} & Line Ratios & 90 & - & $\sim$10000\ensuremath{^a} & \\
 & VII & 25-35\ensuremath{^b} & Line Ratios & 50 & - & $\sim$10000\ensuremath{^a} & \\
 & VIII & 25-35\ensuremath{^b} & Line Ratios & 70 & - & $\sim$10000\ensuremath{^a} & \\
 & IX & 25-35\ensuremath{^b} & Line Ratios & 20 & - & $\sim$10000\ensuremath{^a} & \\
 & X & 25-35\ensuremath{^b} & Line Ratios & 160 & - & $\sim$10000\ensuremath{^a} & \\
 & XI & 25-35\ensuremath{^b} & Line Ratios & 60 & - & $\sim$10000\ensuremath{^a} & \\
 & XII & 25-35\ensuremath{^b} & Line Ratios & 20 & - & $\sim$10000\ensuremath{^a} & \\
 & XIII & 25-35\ensuremath{^b} & Line Ratios & 30 & - & $\sim$10000\ensuremath{^a} & \\
 & XIV & 25-35\ensuremath{^b} & Line Ratios & 70 & - & $\sim$10000\ensuremath{^a} & \\
 & XV & 25-35\ensuremath{^b} & Line Ratios & 110 & - & $\sim$10000\ensuremath{^a} & \\
 & XVI & 25-35\ensuremath{^b} & Line Ratios & 20 & - & $\sim$10000\ensuremath{^a} & \\
 & XVII & 25-35\ensuremath{^b} & Line Ratios & 200 & - & $\sim$10000\ensuremath{^a} & \\ \hline
\caption*{\ensuremath{^a}Assumption/Hypothesis or uncertain value as opposed to sound measurement \\
\ensuremath{^b}Expansion velocity is used as an estimate for the shock velocity \vspace{0.25cm} \\
\hypertarget{references}{References}: $^{(a)}$ \citealt{1991ApJ...366..484B}, $^{(b)}$ \citealt{1983MNRAS.204..273L}, $^{(c)}$ \citealt{2004A&A...426..567M}, $^{(d)}$ \citealt{1983RMxAA...8..155B}, $^{(e)}$ \citealt{1995ApJ...449..681S}, $^{(f)}$ \citealt{2008A&A...481..705B}, $^{(g)}$ \citealt{2002A&A...385.1042B}, $^{(h)}$ \citealt{2009A&A...499..789B}, $^{(i)}$ \citealt{1994ApJ...430..757R}, $^{(j)}$ \citealt{2003A&A...405..591M}, $^{(k)}$ \citealt{2013MNRAS.431..279S}, $^{(l)}$ \citealt{2007MNRAS.381..308B}, $^{(m)}$ \citealt{1980MNRAS.192..731Z}, $^{(n)}$ \citealt{2001A&A...370..265M}, $^{(o)}$ \citealt{1976A&A....49..119S}, $^{(p)}$ \citealt{1983RMxAA...8...59R}, $^{(q)}$ \citealt{1988PASP..100..461B}, $^{(r)}$ \citealt{1991ApJ...373..567L}, $^{(s)}$ \citealt{1993ApJ...405..608W}, $^{(t)}$ \citealt{2005A&A...443..175B}, $^{(u)}$ \citealt{2008Ap&SS.318..207G}, $^{(v)}$ \citealt{2017MNRAS.469..516N}, $^{(w)}$ \citealt{1977ApJ...215L..69G}, $^{(x)}$ \citealt{1981ApJ...250..222R}, $^{(y)}$ \citealt{1985ApJ...292...29F}, $^{z}$ \citealt{2002A&A...388..355M}, $^{(aa)}$ \citealt{2004A&A...424..583B}, $^{(ab)}$ \citealt{1984ApJ...282..161B}, $^{(ac)}$ \citealt{1989MNRAS.237.1109W}, $^{(ad)}$ \citealt{1989ApJ...340..362H}, $^{(ae)}$ \citealt{2001A&A...371..300M}, $^{(af)}$ \citealt{2002A&A...396..225B}, $^{(ag)}$ \citealt{2015ApJ...812...37F}, $^{(ah)}$ \citealt{2003A&A...398..153M}, $^{(ai)}$ \citealt{1992ApJ...400..214L}, $^{(aj)}$ \citealt{1994ApJ...420..721H}, $^{(ak)}$ \citealt{1999ApJ...518..324B}, $^{(al)}$ \citealt{2001AJ....122..938D}, $^{(am)}$ \citealt{2015ApJ...805..152R}, $^{(an)}$ \citealt{2016ApJ...819L..32K}, $^{(ao)}$ \citealt{2014ApJ...791...30M}, $^{(ap)}$ \citealt{1977A&AS...28..439D}, $^{(aq)}$ \citealt{2003A&A...408..237M}, $^{(ar)}$ \citealt{1976A&A....51..159S}, $^{(as)}$ \citealt{2004A&A...415.1051M}, $^{(at)}$ \citealt{2009Ap&SS.324...17G}, $^{(au)}$ \citealt{2007A&A...461..991M}, $^{(av)}$ \citealt{1981Natur.291..132B}, $^{(aw)}$ \citealt{2018MNRAS.473.1705S}, $^{(ax)}$ \citealt{2014MNRAS.441.2996A}, $^{(ay)}$ \citealt{2014ApJ...789..138P}, $^{(az)}$ \citealt{2011ApJ...736..109F}, $^{(ba)}$ \citealt{2002A&A...383.1011M}, $^{(bb)}$ \citealt{2005A&A...435..141M}, $^{(bc)}$ \citealt{1994ApJ...434..635H}, $^{(bd)}$ \citealt{1997AJ....113..767F}, $^{(be)}$ \citealt{2000A&A...353..371M}, $^{(bf)}$ \citealt{1981ApJ...247..148F}, $^{(bg)}$ \citealt{2001ApJ...547..995G}, $^{(bh)}$ \citealt{2017ApJ...846..167K}, $^{(bi)}$ \citealt{1980ApJ...242..592B}, $^{(bj)}$ \citealt{2008ApJS..174..379F}, $^{(bk)}$ \citealt{1983ApJ...270L..53F}, $^{(bl)}$ \citealt{1995AJ....110.2876F}, $^{(bm)}$ \citealt{2007MNRAS.376..929G}, $^{(bn)}$ \citealt{2016ApJ...826..108K}, $^{(bo)}$ \citealt{2010AJ....140.1163F}, $^{(bp)}$ \citealt{1979AuJPh..32..113L}, $^{(bq)}$ \citealt{2018MNRAS.478.1987H}, $^{(br)}$ \citealt{2018RAA....18..111R}, $^{(bs)}$ \citealt{2010AJ....139.2083C}, $^{(bt)}$ \citealt{2013ApJ...765..152L}, $^{(bu)}$ \citealt{1984ApJ...281..658F}, $^{(bv)}$ \citealt{1982RMxAA...5..127R}, $^{(bw)}$ \citealt{2014RMxAA..50..323A}, $^{(bx)}$ \citealt{2012MNRAS.419.1413S}, $^{(by)}$ \citealt{1995ApJ...439..365S}, $^{(bz)}$ \citealt{2000A&A...359..316B}, $^{(ca)}$ \citealt{2003ApJ...589..242S}, $^{(cb)}$ \citealt{1986ApJ...309..667R}, $^{(cc)}$ \citealt{1979MNRAS.188..357G}, $^{(cd)}$ \citealt{1984ApJ...282..135D}, $^{(ce)}$ \citealt{2006AJ....132..360W}, $^{(cf)}$ \citealt{1977MNRAS.181..541L}, $^{(cg)}$ \citealt{1983AJ.....88.1210R}, $^{(ch)}$ \citealt{1996A&A...310L...1B}, $^{(ci)}$ \citealt{2007MNRAS.374.1441S}, $^{(cj)}$ \citealt{1990ApJ...358L..13L}, $^{(ck)}$ \citealt{2003A&A...407..249S}, $^{(cl)}$ \citealt{2013MNRAS.435..910H}, $^{(cm)}$ \citealt{1977ApJ...214..179D}, $^{(cn)}$ \citealt{1980PASP...92..603D}, $^{(co)}$ \citealt{1987ApJ...315L.135K}, $^{(cp)}$ \citealt{2002ApJ...572..888G}, $^{(cq)}$ \citealt{2007ApJ...659.1257R}, $^{(cr)}$ \citealt{2013Sci...340...45N}, $^{(cs)}$ \citealt{2017ApJ...851...12R}, $^{(ct)}$ \citealt{1986MNRAS.222..593M}, $^{(cu)}$ \citealt{1997PASP..109..990C}, $^{(cv)}$ \citealt{2007MNRAS.381..377S}, $^{(cw)}$ \citealt{2001MNRAS.325..287W}, $^{(cx)}$ \citealt{2022MNRAS.515..339P}, $^{(cy)}$ \citealt{2023ApJ...948...97S}, $^{(cz)}$ \citealt{2023ApJ...954...34R}, $^{(da)}$ \citealt{2022MNRAS.516.6055R}, $^{(db)}$ \citealt{2023MNRAS.526.1112D}}
\end{xltabular}

%% file: tables/line_table.tex
\begin{xltabular}{1.34\textwidth}{p{3.2cm}|XXXXp{1.3cm}X|XXp{1.5cm}|p{1.5cm}p{1.5cm}p{1.5cm}}
\caption{Spectral line strengths for different regions within SNRs in our sample. The table is designed to be read horizontally from left to right. Colons preceding numbers indicate uncertain values. This is a small excerpt of the table. The full version can be found \href{https://vizier.cds.unistra.fr/viz-bin/VizieR}{here}.}
\label{tab:lines} \\
\hline\hline
Object \phantom{\Large I} & \multicolumn{6}{c|}{G78.2+2.1} & \multicolumn{3}{c|}{G82.2+5.3} & \multicolumn{3}{c}{G85.9-0.6} \\ \hline
Name(s) & \multicolumn{6}{c|}{DR4, $\gamma$ Cygni SNR} & \multicolumn{3}{c|}{W63} & \multicolumn{3}{c}{-} \\ \hline
Region/Slit Position & Area Ia & Area Ib & Area IIa & Area IIb & Area IIc & Area IId & I (West) & II (East) & III (South) & Area 1 & Area 2 & Area 3 \\ \hline
{[OII]} $\lambda$$\lambda$7320+30 & - & - & - & - & - & - & - & - & - & - & - & - \\
{[OII]} $\lambda$7330 & - & - & - & - & - & - & - & - & - & - & - & - \\
{[OII]} $\lambda$7325 & - & - & - & - & - & - & - & - & - & - & - & - \\
{[OII]} $\lambda$7319 & - & - & - & - & - & - & - & - & - & - & - & - \\
{[Ca II]} $\lambda$7291 & - & - & - & - & - & - & - & - & - & - & - & - \\
{[Fe II]} $\lambda$7155 & - & - & - & - & - & - & - & - & - & - & - & - \\
{[Ar III]} $\lambda$7135 & - & - & - & - & - & - & - & - & - & - & - & - \\
{[SII]} $\lambda$6725 & - & - & - & - & - & - & - & - & - & - & - & - \\
{[SII]} $\lambda$6717+6731 & - & - & - & - & - & - & - & - & - & - & - & - \\
{[SII]} $\lambda$6731 & 31.2 & 11.2 & 12.2 & 11.9 & 13.9 & 16.7 & 25 & 56 & 7.6 & 21 & 21 & 23 \\
{[SII]} $\lambda$6717 & 29.6 & 13.2 & 13.4 & 11.5 & 13.8 & 16.5 & 35 & 78 & 10.8 & 23 & 24 & 27 \\
He I $\lambda$6678 & - & - & - & - & - & - & - & - & 1.3 & - & - & - \\
{[NII]} $\lambda$6583 & 53.7 & 39.4 & 44.6 & 39.7 & 43.3 & 45.7 & 63 & 141 & 32.2 & 33 & 31 & 32 \\
H$\alpha$ $\lambda$6563 & 100 & 100 & 100 & 100 & 100 & 100 & 100 & 100 & 100 & 100 & 100 & 100 \\
{[NII]} $\lambda$6548 & 9.4 & 5 & 12.2 & 11.9 & 11.7 & 12.7 & 15 & 44 & 9.7 & 10 & 11 & 14 \\
{[NII]} $\lambda$6548 + $\lambda$6583 & - & - & - & - & - & - & - & - & - & - & - & - \\
{[OI]} $\lambda$6300 + $\lambda$6363 & - & - & - & - & - & - & - & - & - & - & - & - \\
{[OI]} $\lambda$6364 & - & - & - & - & - & :0.7 & - & 22 & - & - & - & - \\
{[OI]} $\lambda$6300 & :3.8 & - & 2.1 & 0.9 & :0.7 & 2.7 & - & 75 & - & 17 & 13 & 6 \\
He I $\lambda$5876 & - & - & 2.8 & 2.7 & 2.3 & 2 & - & - & - & - & - & - \\
He I $\lambda$5872 & - & - & - & - & - & - & - & - & 3.5 & - & - & - \\
{[N II]} $\lambda$5755 & - & - & - & - & - & - & - & - & - & - & - & - \\
{[N I]} $\lambda$5199 & - & - & - & - & - & - & - & - & - & - & - & - \\
{[OIII]} $\lambda$4959 + $\lambda$5007 & - & - & - & - & - & - & - & - & - & - & - & - \\
{[OIII]} $\lambda$5007 & :4.9 & :8 & 10.2 & 7 & 3.8 & 6.4 & 105 & 511 & 5 & 44 & 35 & 29 \\
{[OIII]} $\lambda$4959 & - & - & :2.6 & 2.1 & :0.9 & :1.8 & 34 & 159 & 1.5 & 15 & 21 & 13 \\
H$\beta$ $\lambda$4861 & 9.8 & 14.4 & 17.9 & 15.8 & 13.4 & 18.5 & 25 & 15 & 11.4 & 52 & 48 & 34 \\
HeII $\lambda$4686 & - & - & - & - & - & - & - & - & - & - & - & - \\
{[OIII]} $\lambda$4363 & - & - & - & - & - & - & - & - & - & - & - & - \\
H$\gamma$ $\lambda$4340 & - & - & - & - & - & - & - & - & - & - & - & - \\
H$\delta$ $\lambda$4102 & - & - & - & - & - & - & - & - & - & - & - & - \\
{[S II]} 4071 & - & - & - & - & - & - & - & - & - & - & - & - \\
{[NeIII]} + H $\lambda$3968 & - & - & - & - & - & - & - & - & - & - & - & - \\
He I + H $\lambda$3889 & - & - & - & - & - & - & - & - & - & - & - & - \\
{[NeIII]} $\lambda$3869 & - & - & - & - & - & - & - & - & - & - & - & - \\
{[NeIII]} $\lambda$3867 & - & - & - & - & - & - & - & - & - & - & - & - \\
{[OII]} $\lambda$3727 & - & - & - & - & - & - & - & - & - & - & - & - \\
Normalization & H$\alpha$ & H$\alpha$ & H$\alpha$ & H$\alpha$ & H$\alpha$ & H$\alpha$ & H$\alpha$ & H$\alpha$ & H$\alpha$ & H$\alpha$ & H$\alpha$ & H$\alpha$ \\
$^a$c & 5.6E-18 & 5.0E-18 & 8.1E-18 & 1.49E-7 & 2.39E-17 & 6.2E-18 & 1.5E-18 & 4.3E-19 & 7.2E-18 & Not given. & Not given. & Not given. \\
Units & \multicolumn{6}{c|}{{erg cm$^{-2}$ s$^{-1}$ arcsec$^{-2}$}} & \multicolumn{3}{c|}{{erg cm$^{-2}$ s$^{-1}$ arcsec$^{-2}$}} & - & - & - \\
$^b$Dereddened? & - & - & - & - & - & - & - & - & - & Y & Y & Y \\
Reference & \multicolumn{6}{c|}{ap} & \multicolumn{3}{c|}{as} & \multicolumn{3}{c}{at} \\ \hline
\caption*{\raggedright \ensuremath{^a}Multiplying coefficient to obtain true intensity. \\
\ensuremath{^b}Whether corrected for interstellar extinction or not. If empty, uncorrected or not clarified. \vspace{0.25cm} \\
The references to the publications we have retrieved our data from are given in \hyperlink{references}{Appendix B}.}
\end{xltabular}